\documentstyle[12pt]{article}   

\large
\newcommand{\beq}{\begin{equation}}
\newcommand{\eeq}{\end{equation}}

\makeatletter
\@addtoreset{equation}{section}
\makeatother

\newtheorem{Theorem}{Theorem}[section]
\newtheorem{Definition}{Definition}[section]
\newtheorem{Lemma}{Lemma}[section]
\newtheorem{Corollary}{Corollary}[section]

\def\be{\begin{equation}}
\def\ee{\end{equation}}
\def\ba{\begin{eqnarray}}
\def\ea{\end{eqnarray}}

\def\ag{{{\cal A}/{\cal G}}}

\def\agb{{\overline {{\cal A}/{\cal G}}}}

\def\Comp{{\mathchoice
{\setbox0=\hbox{$\displaystyle\rm C$}\hbox{\hbox to0pt
{\kern0.4\wd0\vrule height0.9\ht0\hss}\box0}}
{\setbox0=\hbox{$\textstyle\rm C$}\hbox{\hbox to0pt
{\kern0.4\wd0\vrule height0.9\ht0\hss}\box0}}
{\setbox0=\hbox{$\scriptstyle\rm C$}\hbox{\hbox to0pt
{\kern0.4\wd0\vrule height0.9\ht0\hss}\box0}}
{\setbox0=\hbox{$\scriptscriptstyle\rm C$}\hbox{\hbox to0pt
{\kern0.4\wd0\vrule height0.9\ht0\hss}\box0}}}}
\def\Co{{\mathchoice
{\setbox0=\hbox{$\displaystyle\rm C$}\hbox{\hbox to0pt
{\kern0.4\wd0\vrule height0.9\ht0\hss}\box0}}
{\setbox0=\hbox{$\textstyle\rm C$}\hbox{\hbox to0pt
{\kern0.4\wd0\vrule height0.9\ht0\hss}\box0}}
{\setbox0=\hbox{$\scriptstyle\rm C$}\hbox{\hbox to0pt
{\kern0.4\wd0\vrule height0.9\ht0\hss}\box0}}
{\setbox0=\hbox{$\scriptscriptstyle\rm C$}\hbox{\hbox to0pt
{\kern0.4\wd0\vrule height0.9\ht0\hss}\box0}}}}
\def\Rl{{\mathchoice
{\setbox0=\hbox{$\displaystyle\rm R$}\hbox{\hbox to0pt
{\kern0.4\wd0\vrule height0.9\ht0\hss}\box0}}
{\setbox0=\hbox{$\textstyle\rm R$}\hbox{\hbox to0pt
{\kern0.4\wd0\vrule height0.9\ht0\hss}\box0}}
{\setbox0=\hbox{$\scriptstyle\rm R$}\hbox{\hbox to0pt
{\kern0.4\wd0\vrule height0.9\ht0\hss}\box0}}
{\setbox0=\hbox{$\scriptscriptstyle\rm R$}\hbox{\hbox to0pt
{\kern0.4\wd0\vrule height0.9\ht0\hss}\box0}}}}

\title{QSD III :\\ Quantum Constraint Algebra and Physical Scalar Product in
Quantum General Relativity}
\author{T. Thiemann\thanks{thiemann@math.harvard.edu}
\thanks{New Address : Albert-Einstein-Institut,
Max-Planck-Institut f\"ur Gravitationsphysik, Schlaatzweg 1, 14473 
Potsdam, Germany, Internet : thiemann@aei-potsdam.mpg.de} \\
       Physics Department, Harvard University, \\
       Cambridge, MA 02138, USA}
\date{{\small \today\\Preprint HUTMP-97/B-363}}

\begin{document}

\maketitle

\begin{abstract}
This paper deals with several technical issues of non-perturbative 
four-dimensional Lorentzian canonical quantum gravity in the continuum
that arose in connection with the recently constructed Wheeler-DeWitt 
quantum constraint operator.\\
1) The Wheeler-DeWitt constraint mixes the previously 
discussed diffeomorphism superselection sectors which thus become 
spurious,\\ 
2) Thus, the inner product for diffeomorphism invariant states can be fixed
by requiring that diffeomorphism group averaging is a partial isometry,\\
3) The established non-anomalous constraint algebra is clarified by 
computing commutators of duals of constraint operators,\\
4) The full classical constraint algebra is faithfully implemented on 
the diffeomorphism invariant Hilbert space in an appropriate sense,\\ 
5) The Hilbert space of diffeomorphism invariant states can be made
separable if a natural {\em new} superselection principle is 
satisfied,\\ 
6) We propose a natural physical scalar product for quantum general 
relativity by extending the group average approach to the case of 
{\em non-self-adjoint} constraint operators like the Wheeler-DeWitt constraint
and\\ 
7) Equipped with this inner product, the construction of physical 
observables is straightforward.
\end{abstract}

\section{Introduction}

Recently \cite{1a,1b,1c}, a new technique was successfully employed to 
quantize the Wheeler-DeWitt quantum constraint operator of four-dimensional 
Lorentzian canonical gravity in the continuum (see also \cite{Baez}
for a nice and compact non-technical exposition of the essential results 
of these papers). More precisely, the results of these papers show that 
while using all the rigorous kinematical framework that has been 
developed in \cite{AI,AL1,AL2,AL3,MM,almmt}, it is actually possible 
to define a densely defined operator on the underlying kinematical 
Hilbert space whose classical limit corresponds to the classical
Wheeler-DeWitt (or Hamiltonian) constraint of canonical general relativity.

The Hilbert space itself provides a representation of the operator 
algebra in which an $SU(2)$ connection is diagonal, that is, it uses 
very crucially the fact that we are dealing with a connection dynamics 
rather than geometrodynamics formulation of general relativity. The
advantages of the use of a connection as opposed to a three-metric
has been discovered a decade ago by Ashtekar \cite{AA}.
However, in contrast to the original point of view advocated in \cite{AA}
and many later publications
which made essential use of a complex valued connection,
we are dealing with an entirely {\em real} formulation of Lorentzian general 
relativity, a point of view first stressed by Barbero \cite{Barbero}. 
Although the Hamiltonian constraint of general relativity reads much more 
complicated in terms of real than in terms of complex connections,
it allows us to solve the difficult reality conditions which arise  
for the complex variables and which were one of the two major
roadblocks to making progress with these (see, however, \cite{Wick} 
for partial progress in this direction). Moreover, and most importantly,
one is able to rigorously quantize the loop variables introduced by
Gambini and Trias \cite{Gambini} and rediscovered for general 
relativity by Rovelli and Smolin \cite{RS1}, by means of techniques
developed in \cite{AI,AL1,AL2,AL3,MM,almmt}. 

The virtue of \cite{1a,1b,1c}
is that one can overcome the second major roadblock to quantizing the 
Wheeler-DeWitt constraint : its highly non-polynomial structure in terms
of the basic variables which make seem it impossible to give corresponding
quantum analogs any mathematical meaning
(one can make the Wheeler-DeWitt constraint polynomial upon multiplying it 
by a density weighted overall factor but this is not allowed in a 
diffeomorphism invariant theory for reasons explained in \cite{1a,1b} and 
in fact leads to a highly singular operator).
The results of \cite{1a,1b,1c} also show that the operator so constructed 
is anomaly-free in a suitable sense and in \cite{1c} the entire space of 
rigorous solutions to all constraints was constructed. However, the 
quantization programme was still not completed because the following
questions were still unresolved :
\begin{itemize}
\item[1)] In \cite{almmt} the existence of superselection sectors for
diffeomorphism invariant theories of connections was dicovered (theories
which do not have any further constraint). 
The superselection rule holds provided that one can deal with strongly 
diffeomorphism invariant observables only. Is this assumption justified 
in general relativity ?
\item[2)] In \cite{almmt} an infinite number of different inner products,
appropriate for the general solution to the diffeomorphism constraint, 
all of which implement the reality conditions, were constructed. 
An infinite number of overall factors were kept undetermined for the 
inner product on each sector. Does the presence of the Hamiltonian 
constraint somehow fix those factors ? In other words, is there a unique 
group averaging procedure \cite{5,4} (employed there to construct an inner 
product on diffeomorphism invariant states) ?
\item[3)] The operator constructed in \cite{1a,1b} suffers from a huge 
amount of ambiguity : There is not only one operator but an uncountably 
infinite number of them, the freedom being captured by a regularization 
choice. The dual of that operator, determined by evaluating it on a 
{\em diffeomorphism invariant} distribution, however, was shown to be choice
independent because the choice freedom is equivalent to a choice of 
(partial) diffeomorphism gauge fixing. The problem then, however, was
that the dual of the Hamiltonian constraint is not diffeomorphism 
invariant and so multiplying dual operators was not obviously defined. Can 
this be clarified ?
\item[4)] In \cite{1b} the commutator of two Hamiltonian constraint operators
was shown to vanish when evaluated on diffeomorphism invariant states. 
However, we did not compute the commutator directly between duals of these 
operators. More seriously, it is unexpected that the commutator just 
equals zero instead of being proportional to a diffeomorphism constraint 
operator. Finally, we also did not compute the commutator between
the Hamiltonian and the Diffeomorphism constraint operators. 
Can these problems be at least partially solved ? 
\item[5)] The Hilbert space of diffeomorphism invariant states is 
non-separable in the present framework. Can one make it separable by 
imposing a supersection principle ?
\item[6)] In \cite{1c} an algorithm was derived of how to find the 
general solution to all quantum constraints in principle (to find such 
solutions is a matter of computations and can be done in finite time). 
However, we did not provide an inner product on the space of solutions
which would separate a general distribution from a normalizable 
physical state. The main problem is that the Hamiltonian constraint 
is, expectedly, not a self-adjoint operator and so the group averaging
technique which was designed for self-adjoint constraint operators
cannot be applied.
Is it possible to supplement the solution space with an inner product ?
\item[7)] Since in \cite{1c} no physical inner product was derived, 
the issue of Dirac observables was unresolved. Can we construct such 
observables ?
 \end{itemize}
The aim of this paper is to answer all these questions. 
Specifically we show that :
\begin{itemize}
\item[1)] The superselection sectors are spurious : the Hamiltonian 
constraint mixes the sectors and therefore solutions are linear
combinations of vectors from different sectors. Interesting physical 
observables therefore cannot leave these sectors invariant.
\item[2)] Since the superselection sectors are spurious the group 
averaging maps on each sector must come from one big averaging map. 
We show that there is only one such map if one wants to implement the 
group averaging such that strongly diffeomorphism invariant observables 
are self-adjoint and such that a natural class of orthonormal bases on the 
kinematical Hilbert space stays orthonormal after averaging.
\item[3)] As was shown in \cite{1b}, arbitrary products of Hamiltonian 
operators have a dual which is
again choice independent when evaluated on dioffeomorphism invariant states. 
Thus, one can define the product of duals 
to be the dual of the product in reverse order. 
Therefore commutators can be rigorously computed and trivially reproduce 
the results from \cite{1b}.
\item[4)] As foolows immediately from \cite{1b} the commutator algebra 
between dual Hamiltonian 
constraint operators is Abelian which is not expected. However, we 
prove that there exists an operator $\hat{O}(M,N)$ corresponding to $\int 
d^3x (M N_{,a}-M_{,a}N) q^{ab} V_b$
(which is the right hand side of the Poisson bracket $\{H(M),H(N)\}$)
and whose dual is again choice independent and annihilates diffeomorphism 
invariant states. We therefore have the trivial operator identity
$[\hat{H}'(M),\hat{H}'(N)]=-\hat{O}'(M,N)$ on diffeomorphism invariant states
(here the prime stands for dual) and it is in this sense that the Dirac 
algebra is faithfully implemented. 
Morover, the commutator between Hamiltonian and diffeomorphism constraint
closes in the expected way.
\item[5)] We argue that the present results and the structure of the 
Hamiltonian constraint (dynamics) suggest that the Hilbert space 
breaks into a continuous orthogonal sum of isomorphic sectors each of which
is labelled by a countable number of moduli parameters and that no 
physical observable maps between those sectors. Thus, physically one may 
restrict to one of these sectors which turns out to be separable.
\item[6)] We extract the essential features of the group average map for
self-adjoint operators and show that these can be naturally extended to any
non-self-adjoint operator. We show that in quantum mechanical cases our 
proposal does lead to the known or expected result. Astonishingly we find 
that even for self-adjoint operators the group average proposal can
give rise to solutions (obtained via the rigorously defined 
group averaging map) to the constraint which are in fact {\em elements}
of the original Hilbert space, in spite of the fact that the constraint 
operator is unbounded ! We provide quantum mechanics models which mimic
the Euclidean or Lorentzian Hamiltonian constraint and find that all
solutions are normalizable with respect to the original kinematical 
inner product. This suggests that there is nothing wrong with the fact that
the basic solutions found in \cite{1c} are in fact all normalizable
with respect to the inner product obtained by group averaging the 
diffeomorphism constraint. \\
Finally, we use our method to propose a physical inner product for 
quantum general relativity which is very natural. It is very different
from the kinematical diffeomorphism invariant inner product and encodes the 
full dynamics of quantum general relativity.
\item[7)] We take the definition that a Dirac observable maps 
the physical Hilbert space to itself and is self-adjoint thereon.
Given the inner product constructed, the construction of a complete set 
of at least symmetric operators is straightforward.
\end{itemize}
We will leave unresolved the very physical questions of quantum general 
relativity and leave it for future publications. An (incomplete) list
of such questions is given by :\\
1) The problem of time, deparametrization of the theory (see \cite{RR}
for first attempts in this direction).\\
2) Physical interpretation of Dirac observables or how to represent known 
classical observables in the present framework.\\
3) Existence of a semi-classical limit \cite{TTCoh}.\\
4) Black holes and the entropy puzzle from first principles (see 
\cite{R,K,AK} for first success in this direction).\\
5) Connection or relation with string theory.\\
6) Matter coupling and Energy operator \cite{2,2a}.\\
\\
The organization of the article is that we answer the questions raised 
above in the chronological order displayed after introducing some basic
notation at the beginning of the next section.

\section{Diffeomorphism Superselection, Graph Symmetries and Diffeomorphism
Invariant Scalar Product}

The goal of this section is a) to show that there are no superselection 
sectors in quantum gravity once we take the Hamiltonian constraint into
account and b) to select an inner product for diffeomorphism invariant 
states of quantum gravity. On the other hand, the superselection sectors 
arose by 
restricting the algebra of observables to those which strongly commute 
with all diffeomorphisms. We show that if we require that these observables
are to be promoted to self-adjoint operators and that group averaging 
with respect to diffeomorphisms is a partial isometry, then there is
a unique diffeomorphism invariant group averaging inner product.

We begin with a compact review of the relevant notions from \cite{almmt}.
The interested reader is urged to consult this paper and references 
therein, in particular \cite{AI,AL1,AL2,AL3,MM}.\\ 
\\
By $\gamma$ we will denote in the sequel a closed, piecewise analytic graph
embedded into a d-dimensional smooth manifold $\Sigma$ (the case of interest
in general relativity is $d=3$).
The set of its edges will be denoted $E(\gamma)$ and the set of its 
vertices $V(\gamma)$. By suitably subdividing edges into two halves we 
can assume that all of them are outgoing from a vertex (the remaining 
endpoint of the so divided edges are not vertices of the graph because 
they are points of analyticity). Let $A$ be a $G$-connection for a compact 
gauge group
$G$ (the case of interest in general relativity is $G=SU(2)$). We will denote
by $h_e(A)$ the holonomy of $A$ along the edge $e$. Let $\pi_j$ be the 
(once and for all fixed representant of the equivalence class of the 
set of independent) $j$-th irreducible representations of $G$ (in 
general relativity
$j$ is just a spin quantum number) and label each edge $e$ of $\gamma$ 
with a label $j_e$. Let $v$ be an 
$n$-valent vertex of $\gamma$ and let $e_1,..,e_n$ be the edges incident 
at $v$. Consider the decomposition of the tensor product 
$\otimes_{k=1}^n\pi_{j_{e_k}}$ into irreducibles and denote by $c_v$ the
linearly independent projectors onto the singlets that appear.
\begin{Definition} \label{def1}
A spin-network state is defined by
\be \label{1}
T_{\gamma,\vec{j},\vec{c}}(A):=
\mbox{tr}(\otimes_{v\in V(\gamma)}
[c_v\cdot\otimes_{e\in E(\gamma),v\in e}\pi_{j_e}(h_e(A))])
\ee
where $\vec{j}=\{j_e\}_{e\in E(\gamma)},\;\vec{c}=\{c_v\}_{v\in 
V(\gamma)}$. In what follows we will use a compound label
$I\equiv(\gamma(I),\vec{j}(I),\vec{c}(I))$.
\end{Definition}
Thus, a spin-network state is a particular gauge invariant function of 
smooth connections restricted to a graph. Their importance is that they
are an orthonormal basis for a Hilbert space ${\cal H}\equiv{\cal H}_{aux}$, 
called the auxiliary Hilbert space. Orthonormality means that
$$
<T_{\gamma,\vec{j},\vec{c}},T_{\gamma',\vec{j}',\vec{c}'}>_{aux}
=\delta_{\gamma\gamma'}\delta_{\vec{j},\vec{j}'}
\delta_{\vec{c},\vec{c}'}\;.
$$
Another way to describe $\cal H$ is by displaying it as a space of
square integrable functions $L_2(\agb,d\mu_0)$. Here $\agb$ is a space 
of distributional connections modulo gauge transformations, typically 
non-smooth and $\mu_0$ is a 
rigorously defined, $\sigma$-additive, diffeomorphism invariant probability
measure on $\agb$, called the Ashtekar-Lewandowski measure. The space 
$\agb$ is the maximal extension of 
the space $\ag$ of smooth connections modulo gauge transformations such that 
(Gel'fand transforms of) spin-network functions 
are still continuous. The inner product can be extended, with the same
orthonormality relations, to any smooth graph with a finite number 
of edges. It can actually be extended to the full smooth category
\cite{BaSa} but only at the expense of introducing a huge amount of 
additional technicalities. 

We will denote by $\Phi$ the finite linear combinations of spin-network 
functions and call it the space of cylindrical functions. A function 
$f_\gamma$ is said to be cylindrical with respect to a graph 
$\gamma$ whenever it is a finite linear combination of spin-network 
functions on 
that graph such that $\pi_{j_e}$ is not the trivial representation for no
$e\in E(\gamma)$. 
The space $\Phi$ is equipped with some topology, for instance the 
folowing ``Fourier topology" $||f||_1=\sum_I |<T_I,f>|$ (called this 
way for reasons explained in \cite{TTILT}) which 
turns it into a topological vector space. By $\Phi'$ we mean the topological
dual of $\Phi$, that is, the bounded linear functionals on $\Phi$. By the 
Schwarz inequality we have the inclusion $\Phi\subset{\cal H}\subset\Phi'$
(notice, howevver, that this is not strictly speaking a Rigged Hilbert 
Space because the Fourier Topology is not nuclear, see \cite{almmt} for 
another definition).\\ 
This will be enough background for the purpose of the present paper.\\
\\
We now begin entering the diffeomorphism invariant regime.
\begin{Definition} \label{def2}
i) The set of smooth diffeomorphisms which leave a piecewise analytic 
graph $\gamma$ piecewise analytic is denoted by 
$\mbox{Diff}_\gamma(\Sigma)$. The set of all smooth diffeomorphisms 
will be denoted $\mbox{Diff}(\Sigma)$.\\
ii) Let $f_\gamma$ be a function cylindrical with respect to a graph
$\gamma$ and $\varphi\in\mbox{Diff}(\Sigma)$. A unitary 
representation of $\mbox{Diff}(\Sigma)$
is defined by $\hat{U}(\varphi)f_\gamma=f_{\varphi(\gamma)}$.
We define also $\hat{U}(\varphi)T_I=:T_{\varphi\cdot I}$.\\
iii) A srongly diffeomorphism invariant operator $\hat{O}$ on $\cal H$ is a 
self-adjoint operator on $\cal H$ such that 
$\hat{U}(\varphi)\hat{O}=\hat{O}\hat{U}(\varphi)$ for any $\varphi\in
\mbox{Diff}(\Sigma)$.
\end{Definition}
Notice that in contrast to \cite{almmt} we do not work with analytical 
but with smooth diffeomorphisms.\\
The naive idea of how to obtain a diffeomorphism invariant state is to 
take an element of $\Phi$ and to consider its orbit under diffeomorphisms.
More precisely we have the following :\\
Given a cylindrical function $f\in\Phi$ we can uniquely decompose it as
$f=\sum_{i=1}^N f_{\gamma_i}$ where $f_{\gamma_i}$ is a function cylindrical
with respect to the graph $\gamma_i$ and all the graphs in the sum are 
considered to be different. The first guess is to consider the orbit
$\{f\}$ of the state $f$ under diffeomorphisms, that is, all cylindrical 
functions $f'$ such that there is a diffeomorphism $\varphi$ with
$\hat{U}(\varphi)f=f'$ and then to just define $[f]:=\sum_{f'\in\{f\}} f'$.
We immediately see that this does not work :\\
To see this consider the simple state $f=T_I+T_J$ where 
$\gamma(I),\gamma(J)$ are disjoint, say. Then there are infinitely many 
analyticity preserving diffeomorphisms $\varphi_I$ to be considered in the 
orbit 
$\{f\}$ which leave $\gamma(I)$ invariant but not $\gamma(J)$ and vice versa.
As a result, $[f]$ would contain the meaningless term 
$[\sum_{\varphi_I} 1]T_I$. The reason for this difficulty is that each graph
has an infinite-dimensional invariance group of diffeomorphisms. 
Therefore, functions cylindrical with respect to non-coinciding graphs have 
to be averaged separately. Thus we are lead to define 
$$
[f]:=\sum_{i=1}^N\sum_{f_i'\in \{f_{\gamma_i}\}} f_i'=\sum_{i=1}^N 
[f_{\gamma_i}]
$$
where clearly $[f_\gamma]$ only depends on $[\gamma]$, the orbit 
(or generalized knot class) of the graph $\gamma$.\\
However, this still does not quite solve the problem because the orbit 
size of different states cylindrical with respect to the same graph can 
be different. To see what the problem is, consider the following 
instructive example\footnote{This example is due to Don Marolf.} :
Let $\gamma$ be the figure-eight loop (with intersection)  and consider two 
spin-network states
on $\gamma$, $T_1:=T_{\gamma,j_1,j_2,c}$ with $j_1\not=j_2$ and $T_2:=
T_{\gamma,j,j,c'}$ where $c,c'$ are some contractors and $j_1,j_2,j$ are 
certain irreducible representations. Now notice that there is a 
diffeomporphism $\varphi_0$ (the one that rotates the graph by $180$ 
degrees) which leaves $T_2$ invariant but not $T_1$. As a result we see that
the orbit size of $T_1$ is double as large as that of $T_2$. Therefore we 
find that $[T_1+T_2]=[T_1]+2[T_2]$ so that the $[.]$ operation is not at 
all linear !\\
Since we want a linear operation as otherwise we will not preserve the 
linear structure of quantum theory, we must modify the definition of
$[f_\gamma]$ even further, namely, we must choose a basis, write $f_\gamma$
in terms of this basis and average each basis element separately. 
Thus, the group averaging becomes basis dependent, for each basis we get 
another average map.\\
Not every basis is allowed : we want the resulting 
average map to commute with the diffeomorphism group. Notice that this is 
not an entirely trivial requirement because given a basis element $B_I$
and a diffeomorphism $\varphi$ we are supposed to decompose 
$\hat{U}(\varphi)B_I$ into basis elements $B_{I'}$ before averaging 
and then to average the $B_{I'}$. Thus, it is not clear that 
$[\hat{U}(\varphi)B_I]$ is the same as $\hat{U}(\varphi)[B_I]=[B_I]$.
We will call a basis with this property an {\em allowed basis}.\\

This basis dependence is an ugly feature because it
implies that unitary operators are not promoted to unitary operators. 
However, we will see that the Hilbert space splits into two orthogonal
subspaces ${\cal H}^S,{\cal H}^A$ which are basis independently defined :
Namely, ${\cal H}^A$ is nothing else than the kernel of {\em every}
group averaging map while ${\cal H}^S$ is its orthogonal complement which
shows that both spaces are invariantly defined. In each of the subspaces 
the group averaging can be defined {\em basis-independently} while 
being a linear a map. In other words, there are preferred allowed bases of
$\cal H$ which respect the orthogonal split ${\cal H}=
{\cal H}^S\oplus{\cal H}^A$ and the group average with respect to either 
of them is equivalent. As one might expect from the above 
analysis, the spin-network basis is {\em not} a preferred basis, although
it is closely related to one.\\
Let us make precise what we have just said.
\begin{Definition} \label{def3}
i) Let $\{B_I\}_I$ be an orthonormal basis for $\cal H$ (for instance
a spin-network state basis with $B_I=T_I$, without loss of generality we 
can assume that still $I=(\gamma(I),\lambda(I)$). Consider 
the orbit of a 
basis element $B_I$ under $\mbox{Diff}_{\gamma(I)}(\Sigma)$, that is, 
$$
\{B_I\}:=\{\hat{U}(\varphi)B_I,\;\varphi\in\mbox{Diff}_{\gamma(I)}(\Sigma)\}.
$$ 
The group average of $B_I$ is defined to be 
\be \label{2}
B_{[I]}:=[B_I]:=\sum_{B\in\{B_I\}} B\in \Phi'
\ee
and the group average of any $f=\sum_{k=1}^N a_k B_{I_k}\in\Phi$ with 
respect to the basis $\{B_I\}_I$ is defined by $[f]:=\sum_{k=1}^N a_k 
B_{[I_k]}$.\\ 
ii) A basis $\{B_I\}$ is said to be allowed if a) the corresponding 
group averaging commutes with the diffeomorphism group, i.e.
$[\hat{U}(\varphi) B_I]=\hat{U}(\varphi)[B_I]=[B_I]$ for all
$I,\varphi\in\mbox{Diff}_{\gamma(I)}(\Sigma)$ and b) all the vectors
are well-defined distributions, that is, $[B_I]\in\Phi'$ .
In the sequel we will only allow for average maps determined by allowed 
bases and without loss of generality we restrict to orthonormal bases.\\
iii) The diffeomorphism invariant inner product with respect to group 
averaging as determined by an allowed basis $\{B_I\}$ is defined to be
\be \label{3}
<[B_I],[B_J]>_{Diff}:=[B_I](B_J):=\sum_{B\in\{B_I\}}<B,B_J>_{aux}
\ee
and extended by sesquilinearity. We will call the set of group averaged 
cylindrical functions $\Phi'_{Diff}$.
\end{Definition}
Notice that in contrast to \cite{almmt} we do not average with respect to
the {\em graph} $\gamma(I)$ but with respect to the whole {\em state} 
labelled by $I$. This difference will have important consequences with 
respect to the so-called graph-symmetries to which we turn in a moment.
Also notice that we average with respect to analyticity preserving
diffeomorphisms rather than with respect to analytic ones : while 
this reduces the size of diffeomorphism invariant states (it increases 
the orbit and two analytic knot classes may now lie in the same orbit) it 
poses no problem and all the properties derived in \cite{almmt} are 
preserved with the bonus that we do not have to deal with the subtle 
type I and II graphs any longer that were discussed there.

We now construct the subspaces ${\cal H}^S,{\cal H}^A$ mentioned above.
\begin{Definition} \label{def4}
i) Given a graph $\gamma$ the set of graph symmetries of $\gamma$ is the 
subset $G(\gamma)\subset\mbox{Diff}_\gamma(\Sigma)$ of diffeomorphisms
which leave $\gamma$ as a whole invariant but permute its edges and 
vertices.\\ 
ii) Given the orbit $\{T_I\}$ of a spin-network state $T_I$ choose a set
$D(\{T_I\})$ of diffeomorphisms which are in one-to-one correspondence with
the orbit. Let $\varphi_1,\varphi_2\in D(\{T_I\})$ and say that they are 
equivalent iff there is a $\varphi_0\in G(\varphi_1(\gamma(I)))$ such that
$\varphi_2=\varphi_0\circ\varphi_1$. We define $D_0(\{T_I\})$ 
to be the set of equivalence classes of these diffeomorphisms. Clearly
$D_0(\{T_I\})$ is in bijection with $D([\gamma(I)])$ where $[\gamma]$ is the 
orbit of the graph $\gamma$ under $\mbox{Diff}_\gamma(\Sigma)$ and 
$D([\gamma])$ is a set of diffeomorphisms in bijection with $[\gamma]$.\\
iii) Let $T_I$ be a spin-network state. 
Let us write $I=(\gamma(I),\lambda(I))$. Then define 
\be \label{4a}
\Lambda([I]):=\{\lambda,\; \exists \varphi\in G(\gamma(I))\ni
\hat{U}(\varphi)T_I=T_{\gamma(I),\lambda}\}
\mbox{ and } n([I]):=|\Lambda([I])|\;.
\ee
\end{Definition}
To see that $\Lambda{[I]}$ interacts well with the diffeomorphism group 
we need the following lemma.
\begin{Lemma} \label{la1}
The set $\Lambda([I])$ is diffeomorphism invariant.
\end{Lemma}
Proof :\\
Notice that if $\varphi_0(\gamma)=\gamma$ then 
$\varphi\circ\varphi_0\circ\varphi^{-1}(\varphi(\gamma))=
\varphi(\gamma)$ for any $\varphi\in\mbox{Diff}_\gamma(\Sigma)$. 
We conclude 
that $G(\varphi(\gamma))=\varphi\circ G(\gamma) \circ\varphi^{-1}$. 
Let $\varphi'\in D(\{T_I\})$. Then we may pick 
$\varphi\in D_0(\{T_I\})$ such that 
$\hat{U}(\varphi)T_I=T_{\varphi'(\gamma(I)),\lambda(I)}$. Since there is 
$\varphi_0\in G(\varphi'(\gamma(I)))=G(\varphi(\gamma(I)))$ such that
$\varphi'=\varphi_0\circ\varphi$ we see that
$\Lambda([\varphi'\cdot I])=\Lambda([I])$.\\
$\Box$\\
The existence of a non-trivial graph-symmetry group implies that certain 
different labellings of the graph are related by diffeomorphisms. \\
Example :\\
Consider, again, 
the figure-8 loop $\gamma=\alpha_1\circ\alpha_2$. Label  
the loop $\alpha_i$ with $j_i,\;i=1,2$ and the vertex with $c$. The figure-8 
loop has a graph symmetry-group consisting of $\varphi_0,\mbox{id}$ where 
$\varphi_0(\alpha_i)=\alpha_j,\;i\not=j$ and thus $\Lambda([I])
=\{(j_1,j_2,c),(j_2,j_1,c)\}$ provided $j_1\not=j_2$ and $\Lambda([I])
=\{(j,j,c')\}$ if $j_1=j_2=j$. 

We see that the orbit size of the state 
depends not only on the graph but also on the labels. It is for this 
reason that the spin-network states are not a well-adapted basis for the
group-averaging with respect to diffeomorphisms. For example, consider 
a strongly diffeomorphism invariant operator $\hat{O}$ and the states
$\psi_1=T_{\gamma,j_1,j_2,c}, \psi_2=T_{\gamma,j_2,j_1,c},
\psi_3=T_{\gamma,j,j,c'}$ with 
$j_1\not=j_2$ as above with $\hat{U}(\varphi_0)\psi_1=\psi_2,\;
\hat{U}(\varphi_0)\psi_2=\psi_1,\;
\hat{U}(\varphi_0)\psi_3=\psi_3$. First of all we see that
$[\psi_1]=[\psi_2]$ (here we are tentatively assuming that the 
spin-network basis is an allowed basis and average with respect to it). 
Moreover, suppose that $\hat{O}\psi_1=a\psi_1+b\psi_2+c\psi_3$,\\
$\hat{O}\psi_2=b\psi_1+a\psi_2+c\psi_3$,\\
$\hat{O}\psi_3=d\psi_1+d\psi_2+e\psi_3$ with arbitrary complex 
coefficients $a,b,c,d,e$.\\
This is the most general strongly diffeomorphism invariant operator on 
these three states. If it is self-adjoint then $a,b,e$ are real and 
$\bar{c}=d$. Now let us see whether with the present definition 
$\hat{O}$ is promoted to a self-adjoint operator on ${\cal H}_{Diff}$.
The necessary and sufficient condition is that 
$[\psi_i](\hat{O}\psi_j)=[\hat{O}\psi_i](\psi_j)$. But
$[\psi_1](\hat{O}\psi_3)=2d$ because $[\psi_1]=[\psi_2]$ so that the 
orbit of $\psi_1$ also contains $\psi_2$ while 
$[\hat{O}\psi_1](\psi_3)=\bar{c}=d$ contradicting symmetry. \\
On the other hand, this example shows how to cure the situation by a 
suitable change of basis :
consider the orthonormal states $\psi_\pm:=(\psi_1\pm\psi_2)/\sqrt{2}$.
Then by definition $[\psi_+]=\sqrt{2}[\psi_1],\;[\psi_-]=0$, that is,
the group average as defined by (\ref{2}) has a huge kernel even when 
restricted to vectors defined on the same graph. Notice that 
group averaged states remain orthonormal if not zero, provided we divide 
the average of $\psi_+$ by the number of spin-network states involved 
(here two) since they come out of the averaging anyway :\\
$\frac{1}{2}[\psi_+](\psi_+)=\frac{\sqrt{2}}{2\sqrt{2}}
[\psi_1](\psi_1+\psi_2)=1$ (still averaging with respect to the 
spin-network basis) and the so
defined inner product is hermitean. In terms of the new vectors the operator
$\hat{O}$ becomes \\
$\hat{O}\psi_+=\frac{a+b}{\sqrt{2}}\psi_+ +\sqrt{2}c\psi_3$,\\
$\hat{O}\psi_-=\frac{a-b}{\sqrt{2}}\psi_-$,\\
$\hat{O}\psi_3=d\sqrt{2}\psi_+ +e\psi_3$ and we see that it leaves the kernel
of the group averaging map {\em invariant} and moreover does not have any
matrix elements between the kernel and its complement. 
We now verify that 
$\frac{1}{2}[\psi_+](\hat{O}\psi_3)=
\frac{d}{\sqrt{2}}=\frac{\bar{c}}{\sqrt{2}}=\frac{1}{2}
[\hat{O}\psi_+](\psi_3)$
is indeed self-adjoint. How did this happen ? 
It happened because the 
group average was defined with respect to the spin-network basis and so 
one must expect a basis dependence of the resulting diffeomorphism 
invariant inner product. Indeed, the average with respect to the new basis 
$\psi_\pm$ 
modified the averaging as compared to the spin-network basis and in fact, 
the spin-network basis is not 
well-adapted to difeomorphism invariance because the group average of two
orthogonal states with the same graph on $\cal H$ can result in 
diffeomorphism invariant states which are identical and so have unit 
inner product. By changing to the basis involving $\psi_\pm$ we have done 
two things : first, the symmetric state $\psi_+$ is essentially what is 
important after group averaging while the anti-symmetric state $\psi_-$ is 
unimportant and becomes zero after averaging. Morover, the 
representation of diffeomorphism 
invariant self-adjoint operators is completely reducible with symmetric 
and anti-symmetric states as invariant subspaces.

We will now show that these features of the example are true in general
and that the generalization of the basis of the $\psi_\pm$ is preferred.
\begin{Definition} \label{def5}
i) Given a spin-network state $T_I$, label the $n=n([I])$ states 
$T_{\gamma(I),\lambda_k}:=T_{I,k}$ where $\{\lambda_1,..,\lambda_n\}
=\lambda([I])$. Now construct the orthonormal states
\ba \label{4}
T^S_I&:=&\frac{1}{\sqrt{n}}\sum_{k=1}^n T_{I,k}\nonumber\\
T^A_{I,k}&:=&\frac{1}{\sqrt{k(k+1)}}[\sum_{l=1}^k T_{I,l}-k T_{I,k+1}]
\mbox{ for }k=1,..,n-1\;.
\ea
ii) Let ${\cal H}^S$ be (the completion of the) span of the $T^S_I$ and 
likewise for ${\cal H}^A$. Then ${\cal H}={\cal H}^S\oplus{\cal H}^A$. 
${\cal H}^A$ is called the kernel of the average map [for reasons 
explained in the next theorem].\\
iii) A basis such that each basis vector lies either in ${\cal H}^S$ or 
in ${\cal H}^A$ is said to be adapted.
\end{Definition}
That these states are orthonormal is readily verified by induction over
$k=1,..,n$. We see that we arrive at a new orthonormal decomposition of 
the Hilbert space of states and of the space of functions 
cylindrical with respect to given graph $\gamma$.
\begin{Theorem} \label{th1}
i) The orbit size of any state of the form $T^S_I$ is given by
$D([\gamma(I)])$, that is, it only depends on the orbit size of the graph
and not on the state.\\
ii) The adapted basis given by the $T^S_I$ and $T^A_{I,k}$ is allowed.\\
iii) The space $[{\cal H}^A]_\gamma$ is the kernel of every average map 
restricted to functions cylindrical with respect to a graph $\gamma$ for 
any $\gamma$ and therefore is invariantly defined.\\
iv) Two average maps defined by adapted bases of ${\cal H}$ are identical.
We will call the resulting average map $\eta_{Diff}$.
\end{Theorem}
Proof :\\
i)\\
Since the $T^S_I$ are invariant under the graph symmetry group $G(\gamma(I))$
they actually do not depend on $\lambda(I)$ but only on $\Lambda([I])$.
Therfore, the orbit size of $T^S_I$ is that of the set $D_0(\{T_I\})$ 
which in turn is in bijection with $D([\gamma(I)])$. Notice that therefore
$\eta_{Diff}T^S_I=[T_{I,1}]/\sqrt{n([I])}$ where $[T_{I,1}]$ is the average 
with respect to the spin-network baisis. This shows that 
\ba \label{5}
(\eta_{Diff}T^S_I)(T^S_J)
&=&\frac{1}{\sqrt{n([I])n([J])}}[T_{I,1}](T_{J,1}+..+T_{J,n([J])})
\nonumber\\
&=&\frac{n([J])}{\sqrt{n([I])n([J])}}[T_{I,1}](T_{J,1})
=\frac{n([J])}{\sqrt{n([I])n([J])}}\delta_{[\gamma(I)][\gamma(J)]}
\delta_{\Lambda([I])\Lambda([J])}\nonumber\\
&=:&\delta_{[I][J]}
\ea
so the states $\eta_{Diff}T^S_I$ are orthonormalized under the associated 
diffeomorphism invariant inner product. \\ 
ii)\\
We need to show that $[T^S_I]=[\hat{U}(\varphi)T^S_I]$ and
$[T^A_{I,k}]=0=[\hat{U}(\varphi)T^A_{I,k}]$ for each $\varphi,I,k$.
However, by definition, each of the $T^S_I,T^A_{I,k}$ is nothing else 
than a particular linear combination of spin-network states $T_I$
which have the nice feature that a diffeomorphism just can be translated
into an action on the label $I$ : $\hat{U}(\varphi)T_I=
T_{\varphi(\gamma(I)),\lambda(I)}=T_{\varphi\cdot I}$. It follows that 
$\hat{U}(\varphi)T^S_I=T^S_{\varphi\cdot I}$ and 
$\hat{U}(\varphi)T^A_{I,k}=T^A_{\varphi\cdot I,\varphi\cdot k}$.
Thus the result follows from the fact that $T^S_I$ only depends 
on $\Lambda(I)$ which is diffeomorphism invariant and from the fact that
$[T^A_{I,k}]=0$ for any $I,k$.\\ 
iii)\\
$\Rightarrow$\\
First we show that every state of the form $T_I-\hat{U}(\varphi)T_I$,
where $\varphi\in G(\gamma(I))$, is averaged to zero by every averaging 
map, that is, by the averaging map determined by any allowed basis 
$\{B_I\}$. Because the averaging map is linear and these states 
span ${\cal H}^A$, we will have shown that ${\cal H}^A_\gamma$ is in the
kernel of every average map on ${\cal H}^A_\gamma$. \\
Since $T_I$ and $B_I$ belong orthonormal bases there is a unitary operator
$\hat{V}$ with matrix elements $V_{IJ}$ such that $T_I=\sum_J V_{IJ} B_J$
and the sum extends only over labels such that $\gamma(J)=\gamma(I)$.
Now the state $\hat{U}(\varphi)B_J$ is a linear combination of states 
$B_{J'}$ : in other words $\hat{U}(\varphi)B_J=\sum_{J'}L_{JJ'} B_{J'}$. 
Therefore by definition $[\hat{U}(\varphi)B_J]:=\sum_{J'}L_{JJ'}[B_{J'}]
=[B_J]$ because $\{B_I\}_I$ is an allowed basis. We conclude that 
$[T_I]=[\hat{U}(\varphi)T_I]$.\\
$\Leftarrow$\\
Conversely, we need to show that every vector that is averaged to zero 
and that is cylindrical with respect to the same graph $\gamma$ lies in 
${\cal H}^A_\gamma$. Since we have shown already that each 
vector in ${\cal H}^A_\gamma$ has this property we need to show that no 
vector in
${\cal H}^S_\gamma$ has this property because the average map is linear.\\
Let us introduce a new notation : Let $T^S_I:=T^S_i$ and $T^A_{I,k}:=T^A_\mu$
and consider any allowed orthonormal basis $B_I$. There exists a unitary 
operator $\hat{V}$ such that 
$$
T^S_i=\sum_I V_{iI} B_I,\;T^A_\mu=\sum_I V_{\mu I} B_I\mbox{ and }
B_I=\sum_i \bar{V}_{iI}T^S_i+\sum_\mu \bar{V}_{\mu I} T^A_\mu\;.
$$
According to i) all vectors $T^S_i$ are invariant under the graph symmetry
group $G(\gamma)$ while the vectors $T^A_\mu$ are not. In fact, by 
definition of the graph symmetry group, their orbit size is an integral 
multiple
of the orbit size of $\gamma$ which is the orbit size of any $T^S_i$.
Therefore the orbit size of the state $B_I$ is an integral multiple
$n(I)\ge 1$ of the orbit size of $\gamma$ the value of which 
depends on which and how many of the states $T^A_\mu$ occur in its 
decomposition. Combining this with the first half of the proof of 
iii) we find $[B_I]=n(I)\sum_i\bar{V}_{iI} \eta_{Diff} T^S_i$.
Certainly, 
$n(I)$ is a bounded function of $I$ as otherwise not all the $[B_I]$
would be in $\Phi'$ (since $[\eta_{Diff}T^S_I](T^S_J)=
\delta_{\Lambda([I])\Lambda([J])}$).\\
We now ask whether there exist complex numbers $a_i$ such that the vector
$\psi^S:=\sum_i a_i T^S_i\in {\cal H}^S$ is averaged to zero. By 
definition we have 
\ba
[\psi^S]=\sum_i a_i\sum_I V_{iI}[B_I]
&=&\sum_i a_i\sum_I V_{iI}\sum_j n(I)\bar{V}_{jI} \eta_{Diff}T^S_j\nonumber\\
&=&\sum_j \eta_{Diff}T^S_j \sum_i a_i [\sum_I V_{iI}n(I)\bar{V}_{jI}]=0\;
\nonumber\ea
which has a non-trivial solution $\{a_i\}_i$ if and only if the
operator $\hat{M}$ on ${\cal H}^^S_\gamma$ with matrix elements between 
the $T^S_i$
$$
M_{ij}=\sum_I V_{iI}n(I)\bar{V}_{jI}=[\hat{V}\hat{D}\hat{V}^\dagger]_{ij}
$$
is singular where we have defined the operator $\hat{D}$ on $\cal H$ defined 
by $\hat{D}B_I=n(I)B_I$, that is, it has matrix elements $D_{IJ}=
n(I)\delta_{IJ}$ between $B_I,B_J$. Notice that the operators $\hat{D},
\hat{V}$ are bounded so that the resummation that we performed above is 
justified.\\
Since $n(I)\ge 1$, the operator $\hat{D}$ has a square root $\hat{R}$ and 
thus $\hat{M}=[\hat{V}\hat{R}][\hat{V}\hat{R}]^\dagger$. Thus we are asking
whether the projection to ${\cal H}^S$ of a positive definite operator 
on $\cal H$ is singular which is impossible.\\
This shows that $[\psi^S]=0$ if and only if $\psi^S=0$.\\
iv)\\
Let $\{B_I\}_I$ be an allowed basis adapted to the orthogonal decomposition
${\cal H}^S\oplus{\cal H}^A$. That is, each $B_I$ lies entirely only in 
one of these spaces.
Since the $B_I$ define an orthonormal basis there is a 
unitary map $V^S\oplus V^A$ such that $(V^S)^\dagger V^A=(V^A)^\dagger 
V^S=0,\;(V^S)^\dagger V^S=\mbox{id}_S,\;(V^A)^\dagger V^A=\mbox{id}_A$ and
$T^S_I=\sum_J V^S_{IJ} B_J, T^A_\mu =\sum_J V^A_{\mu J} B_J$ where the sum 
runs only over labels $J$ such that $\gamma(I)=\gamma(J)$ (we have 
written $T^A_{I,k}=T^A_\mu$) and also only $B_J$ are involved which are 
either in ${\cal H}^S$ or ${\cal H}^A$ respectively. 
By definition 
$$
[T^S_I]=
\sum_J V^S_{IJ}\sum_{B_J'\in\{B_J\}} B_J'=
\sum_J V_{IJ}\sum_{\varphi\in D(\{B_J\})} \hat{U}(\varphi) B_J
$$
where, as before, $D(\{B_J\})$ is a set of diffeomorphisms which are in 
one to one correspondence with the orbit of $B_J$. Notice that 
$B_J=\sum_K (V^S)^{-1}_{JK} T^S_K$ since
the basis is adapted (there are no $T^A_\mu$ on the right hand 
side). According to i)
the orbit size of all the $T^S_K$ is identical and in fact coincides 
with the orbit size of the graph $\gamma(I)$.
That means that each $\varphi\in
D(\{B_J\})$ which moves $B_J$ at all must move all the $T^S_K$ which 
appear on the right hand side of 
$B_J=\sum_K (V^S)^{-1}_{JK} T^S_K$. It follows that the orbit size of 
$B_J$ is the same as that of $[\gamma(I)]$ as well. 
It follows that $[B_J]=\sum_K (V^S)^{-1}_{JK} \eta_{Diff}T^S_K$
so that, putting these results together, we find
$$
[T^S_I]=\eta_{Diff} T^S_I
$$
where $\eta_{Diff}$ is the average map defined by the basis $T^S_I$.\\
$\Box$\\
Let us verify that $\eta_{Diff} T^S_I\in \Phi'$. If $f\in\Phi$, then 
\ba \label{demons}
|[T^S_I](f)|& :=& \frac{1}{\sqrt{n(I)}}|\sum_{T\in\{T_I\}}<T,f>_{aux}|\le 
\frac{1}{\sqrt{n(I)}} \sum_{T\in\{T_I\}}|<T,f>_{aux}|
\nonumber\\
&\le&
\frac{1}{\sqrt{n(I)}}\sum_J|<T_J,f>_{aux}|=\frac{||f||_1}{\sqrt{n(I)}}
\ea
which shows that  $\eta_{Diff} T^S\in \Phi'_{Diff}$.\\
\\
Let us summarize the contents of Theorem (\ref{th1}) with regard to the
uniqueness of the average map :\\ 
Every average map has the same kernel ${\cal H}^A$ so that the kernel is 
defined independently
of the allowed basis that was used to define the average map. Therefore
the spaces ${\cal H}^S,{\cal H}^A$ are invariantly defined.\\
Being invariantly defined one can consider all bases that are adapted to 
the associated orthogonal decomposition which is natural because the 
space ${\cal H}^A$ does not play any role after moding out with 
respect to the diffeomorphism group. Moreover, as we will see below, the 
subalgebra of the observable algebra given by strongly diffeomorphism 
invariant observables leaves the subspaces ${\cal H}^S,{\cal H}^A$ invariant
so that we have an additional motivation to consider only averages with 
respect to adapted bases in the sequel.\\
Then, when restricting to adapted bases it turns out that there is actually
only one averaging map. \\
Thus, among all the possible averagings, one of them, $\eta_{Diff}$, is 
singled out.\\
\\
We could be content with this result, however, we now ask if we can do even
better : Recall that we defined an average map by first decomposing a 
cylindrical function into components each of which was cylindrical with 
respect to a certain graph. Now diffeomorphic graphs certainly have 
identical orbit sizes but non-diffeomorphic ones do not have, in general,
the same orbit size and as we have seen above, in general their orbit sizes
are ``infinitely different" in the sense that one gets in general not
an element of $\Phi'$ if summing over all vectors in the orbit of a 
vector which is a linear combination of vectors cylindrical with respect to
non-diffeomorphic graphs. It seems therefore that we do not have any 
justification for a universal average map $\eta_{Diff}$, that is, for 
each diffeomorphism equivalence class of graphs $[\gamma]$ we could have a 
different average map $\eta_{Diff}^{[\gamma]}$. Acoording to the above 
analysis, the difference between $\eta_{Diff}$ and $\eta_{Diff}^{[\gamma]}$
can really be only a pre-factor if we choose the canonical way of averaging
and the prefactor is a positive real number which, heuristically 
speaking, accounts for the fact that 
each graph has different orbit size as compared to the orbit size of the 
``universal graph" which contains all graphs (and which therefore has the 
largest possible orbit size) and arises by somehow renormalizing the 
arising infinite pre-factor.\\
The justification for the prefactor in $\eta_{Diff}^{[\gamma]}=a([\gamma])
\eta_{Diff}$ is even more serious given the superselection principle
derived in \cite{almmt}.
Let us shortly review how strongly diffeomorphism invariant 
observables define a superselection rule :
Assume that $\gamma(I)$ and $\gamma(J)$ are 
different graphs and that $\hat{O}$ is a strongly diffeomorphism 
invariant observable which is densely defined on $\Phi$. There are an at 
least countable number of diffeomorphisms $\varphi_n$ which leave 
$\gamma(I)$ invariant but such that $\varphi_n^{-1}(\gamma(J))$ are mutually
different graphs for each $n$. It follows that 
\ba
O_{IJ}&:=&<T_I,\hat{O} T_J>_{aux}=
<\hat{U}(\varphi_n)T_I,\hat{O} T_J>_{aux}=
<T_I,\hat{U}(\varphi_n^{-1})\hat{O} T_J>_{aux}\nonumber\\ &=&
<T_I,\hat{O}\hat{U}(\varphi_n^{-1}) T_J>_{aux}\nonumber\\
\ea
and so $\hat{O}$ has the same matrix element $O_{IJ}$ between $T_I$ and 
an infinite number of mutually orthonormal states. Thus, since $\hat{O}$
was densely defined we conclude that $O_{IJ}=0$. This is the 
superselection rule : strongly diffeomorphism invariant observables 
cannot map between cylindrical functions defined on different graphs. 
In particular, they cannot map between graphs which are non-diffeomorphic.
Therefore, the inner product between diffeomorphism invariant states 
which arise as group averages from different diffeomorphism invariant sectors
are completely unrelated and we may therefore choose each of them separately.

Of course, the existence of the superselection rule rests on the 
assumption that strongly diffeomorphism invariant observables are the only
ones that givve rise to physical observables inquantum gravity.
Now, from \cite{1a,1b} we know that the Hamiltonian constraint operator 
does not
leave the superselected sectors invariant. In particular, the space of 
solutions consists of distributions which are linear combinations of 
diffeomorphism invariant states from different superselected sectors. It 
follows that any non-trivial physical observable, that is, one that maps the 
solution space to itself, and which differs from the identity operator 
mixes different super-selected sectors. Clearly, such an observable 
cannot be strongly diffeomorphism invariant. Strongly diffeomorphism 
invariant operators are built purely from electric field operators (as 
they cannot change the graph) while weakly diffeomorphism invariant 
operators will also depend on the connection through loops. We conclude that 
the restriction to strongly diffeomorphism invariant observables was too 
restrictive and that the superselection sectors are therefore spurious.

Still there remains the issue which value of the constants $a([\gamma])$ 
is the correct one. Is there is a physical motivation for $a_{[I]}=1$ ?
\begin{Theorem} \label{th2}
i) The representation of the algebra of strongly diffeomorphism invariant
obeservables on $\cal H$ is completely reducible with 
${\cal H}^S,{\cal H}^A$ as invariant subspaces.\\
ii) The inner product 
defined by 
$<\eta_{Diff}f,\eta_{Diff}g>_{Diff}:=(\eta_{Diff} f)(g)$ for all
$f,g\in\Phi$, that is, $a([\gamma])=1$, is uniquely 
selected by requiring that a) strongly diffeomorphism invariant 
self-adjoint operators
on $\cal H$ are promoted to self-adjoint operators on ${\cal H}_{Diff}$
and b) $<f,g>_{aux}=<\eta_{Diff} f,\eta_{Diff}g>_{Diff}$ for all $f,g\in 
({\cal H}^S\cap\Phi)_\gamma$. 
\end{Theorem}
Proof :\\
i) \\
Notice that each element $\varphi_0\in G(\gamma(I))$ permutes the 
edges and vertices of $\gamma(I)$ in some particular way. This 
permutation induces a permutation $\sigma_0$ of the elements of the set
$\Lambda([I])$. Thus, $\sigma_0\in S(n([I]))$ where $S(n)$ denotes the 
symmetric group of $n$ elements (the set of these permutations does not 
necessarily contain all elements of $S(n)$ but is a subgroup thereof). The 
vectors $T^S_I$ are therefore invariant 
under $G(\gamma([I]))$ because they are completely symmetric under any 
element of $S(n([I]))$. 

We conclude that for a strongly diffeomorphism 
invariant operator $\hat{O}$ the vector $\hat{O}T^S_I$ is again invariant
under $G(\gamma(I))$ (recall that a strongly diffemorphism invariant operator
must map functions cylindrical with respect to a graph to functions 
cylindrical with respect to the same graph). Next, notice that the 
graph symmetry group acts transitively and fixpoint freely on the set of 
vectors 
$\{T_{I,k}\}_{k=1}^{n([I])}$ by its very definition (every vector can be  
mapped ino any other). It follows that the 
induced subgroup $S'(n([I]))$ of the permutation group on $\Lambda([I])$ 
contains a cyclic element $\sigma$ of order $n([I])$ (that is, one with 
$\sigma^{n([I])}= \mbox{id}$ and $n=n([I])$ is the smallest integer with 
this property). 

To see this just decompose every $\sigma\in S'(n([I]))$ 
into cycles. Take one element which contains a cycle of length $k$
and compose it with another element which contains a cycle of length $l$ 
that involves at least one element which is not involved in the first 
cycle and at least one that is involved in the first cycle. Such an element 
must exist because the graph symmetry group acts 
transitively on $\Lambda([I])$. The result is an element of $S'(n([I]))$
which contains a cycle of length $k+1$ at least. Proceeding inductively, 
the assertion follows.

Given such a cyclic element $\sigma_0\in S'(n([I]))$ corresponding to
$\varphi_0\in G(\gamma(I))$, 
let us label the states $T_{k,I}$ in such a way that 
$\hat{U}(\varphi_0)^k T_{I,1}=T_{I,k+1}$. Let us 
consider a new (non-orthonormal) basis of 
$\mbox{span}\{T^A_{I,k}\}_{k=1}^{n([I])-1}$ 
given by $A_{I,k}:=T_{I,k}-T_{I,k+1},\;k=1,..,n([I])-1$. That indeed every 
$T^A_{I,k}$ is 
a linear combination of these new vectors follows from the identies
$\sum_{l=1}^k T_{I,l}-k T_{I,k+1}=\sum_{l=1}^k[T_{I,l}-T_{I,k+1}]$ and 
$T_{I,l}-T_{I,k+1}=A_{I,l}+..+A_{I,k}$. The virtue of this new basis is that
it allows us to readily verify that 
$$
\hat{\eta}_I A_{I,l}:=(\sum_{k=0}^{n([I])-1} 
\hat{U}(\varphi_0)^k)A_{I,l}=0\;\forall\; l=1,..,n([I])-1
$$ 
since $\hat{U}(\varphi_0)A_{I,l}=A_{I,l+1}$ with $T_{I,n([I])+1}:=
T_{I,1}$ and because $\sum_{k=1}^{(n([I])}A_{I,k}=0$.
We conclude that $\eta_I T^A_{I,k}=0$ for all $k=1,..,n([I])$. \\
We will show that a strongly diffeomorphism invariant operator $\hat{O}$ 
must leave the cylindrical subspaces of ${\cal H}^S$ and ${\cal H}^A$ 
invariant. To that end let be given arbitrary elements 
$f$ of ${\cal H}^S\cap\Phi$ and $g$ of ${\cal H}^A\cap\Phi$. This means
that they are finite linear combinations of the form 
$f=\sum_{i=1}^M s_i T^S_{I_i}$ and $g=\sum_{j=1}^N a_j T^A_{J_j,k_j}$ where 
$k_j\in\{1,..,n([J_j])-1\}$. We need to show that $<f,\hat{O}g>
=<g,\hat{O}f>=0$ for all $f\in{\cal H}^S\cap\Phi,\;g\in{\cal H}^A\cap\Phi$. 
This is equivalent to showing that $<T^S_I,\hat{O}T^A_{J,k}>=
<T^A_{J,k},\hat{O}T^S_I>=0$ for all $I,J$ and $k=1,..,n([J])-1$. Since 
a strongly diffeomorphism invariant operator leaves the graph of 
a spin-network state invariant these equalities follow trivially unless
$\gamma(I)=\gamma(J)$ so let us assume that this is the case.  
We now use the fact proved above that $T^S_I$ is invariant under 
$G(\gamma(I))$. This implies that, since $\gamma(I)=\gamma(J)$,
$\hat{\eta}_J^\dagger T^S_I=T^S_I$. Since by assumption
$[\hat{\eta}_J^\dagger,\hat{O}]=0$ we have
$$
<T^S_I,\hat{O}T^A_{J,k}>=
<\hat{\eta}_J^\dagger T^S_I,\hat{O}T^A_{J,k}>=
<T^S_I,\hat{O}\hat{\eta}_J T^A_{J,k}>=0
$$ 
by construction of $\hat{\eta}_J$. Similarily
$$
<T^A_{J,k},\hat{O}T^S_I>=
<T^A_{J,k},\hat{O}\hat{\eta}_J^\dagger T^S_I>=
<\hat{\eta}_J T^A_{J,k},\hat{O}T^S_I>=0\;.
$$
As an aside we notice that $G(\gamma(I))$ leaves also ${\cal H}^A$ invariant
because the identity operator is obviously strongly diffeomorphism 
invariant.\\
ii)
The most general group averaging with respect to adapted bases is given by 
$\eta_{Diff}^{(a)}:=a([\gamma(I)])\eta_{Diff}$ 
and the associated most general inner product is given by 
$<\eta_{Diff}^{(a)}f,\eta_{Diff}^{(a)}g>_{Diff}:=
(\eta_{Diff}^{(a)}f)(g)$ where $a([\gamma(I)])>0$ depends only on 
the orbit of the graph.
Let us see which 
restrictions are imposed by requiring properties a), b) stated in part
ii) of the theorem.\\
We begin with b). We have already seen in Theorem (\ref{th1}), i) that the 
$\eta_{Diff}T^S_I$
are orthonormal with respect to the inner product given by $a([\gamma])=1$
for all $[\gamma]$. It follows that for any $f,g\in {\cal H}^S_\gamma$ we 
have with $f=\sum_I f_I T^S_I,\;g=\sum_I g_I T^S_I$
$$
<\eta^{(a)}_{Diff}f,\eta^{(a)}_{Diff}g>_{Diff}=\sum_{I,J}\bar{f}_I g_J 
(\eta^{(a)}_{Diff} T^S_I)(T^S_J)=a([\gamma])\sum_I \bar{f}_I g_I=
a([\gamma])<f,g>_{aux}
$$
which fixes $a([\gamma])=1$.\\
We turn to a). Since a strongly diffeomorphism invariant 
operator $\hat{O}$ leaves ${\cal H}^A$ invariant the only non-trivial
matrix elements involving $T^A_{I,k}$ are given by 
$(\eta_{Diff}T^A_{I,k})(\hat{O}T^A_{J,l})=0$ since 
$\eta_{Diff}T^A_{I,k}=0$ and 
$\eta_{Diff}(\hat{O} T^A_{I,k})(T^A_{J,l})=0$ since 
$\eta_{Diff}\hat{O} T^A_{I,k}=0$ because the averaged vector lies in 
${\cal H}^A$ again. So on ${\cal H}^A$ the operator $\hat{O}$ is trivially 
self-adjoint because after averaging it is the zero operator. Thus we 
need to check only whether or not it is true that 
\be \label{6}
<\eta_{Diff}\hat{O}f,\eta_{Diff}g>_{Diff}=
<\eta_{Diff}f,\eta_{Diff}\hat{O}g>_{Diff}\;
\ee
for any $f,g\in {\cal H}^S$. However, this is trivial : Since $\hat{O}$ 
cannot map between functions which are cylindrical with respect to 
diffeomorphism inequivalent graphs we may assume that $f,g$ are in fact
cylindrical with respect to the same graph $\gamma$. Since both lie in 
${\cal H}^S$ and $\hat{O}$ leaves ${\cal H}^S$ invariant 
all vectors in (\ref{6}) have the orbit 
size of $\gamma$. Thus, since $\hat{O}$ commutes with all diffeomorphisms 
and is self-adjoint on ${\cal H}_{aux}$
\ba \label{7}
<\eta_{Diff}\hat{O}f,\eta_{Diff}g>_{Diff}
&=&\sum_{\varphi\in D([\gamma])} <\hat{U}(\varphi)\hat{O}f,g>\nonumber\\
&=&\sum_{\varphi\in D([\gamma])} <\hat{O}\hat{U}(\varphi)f,g>\nonumber\\
&=&\sum_{\varphi\in D([\gamma])} <\hat{U}(\varphi)f,\hat{O}g>\nonumber\\
=<\eta_{Diff}f,\eta_{Diff}\hat{O}g>_{Diff}
\ea
which furnishes the proof.\\
$\Box$\\
Notice that it was crucial in the proof for the satisfaction of the 
reality conditions that strongly diffeomorphism invariant observables 
leave the spaces ${\cal H}^S,{\cal H}^A$ invariant. Interestingly,
the Hamiltonian constraint operator defined in \cite{1a,1b,1c} maps the 
space ${\cal H}^S$ also into itself because it treats all parts of the graph 
democratically. 

The condition that fixes the value $a([\gamma])=1$ is that inner products
between vectors on ${\cal H}^S_\gamma$ are preserved under group averaging.
This implies that the $\eta_{Diff}T^S_I=[T_I]/\sqrt{n([I])}$ are a natural 
orthonormal basis on ${\cal H}_{Diff}$. What is the meaning of this 
condition ? It means that the group averaging defined is a partial 
isometry from ${\cal H}_{aux}$ to ${\cal H}_{Diff}$ (partial because 
$\eta_{Diff}$ has a huge kernel even on ${\cal H}^S$). We can interprete 
this as saying that a reality condition selected the inner product because
for a transformation to be a partial isometry means that it satisfies a
(partial) adjointness condition with respect to the inner product to be 
determined.

\section{The dual Constraint Algebra}

Recall that the action of the Hamiltonian $\hat{H}(N)$ constraint on $\Phi$ 
was 
inherently equipped with a huge amount of ambiguity which had to do with
the precise position of a particular edge to be attached to a graph.
In other words, there is a certain infinite-dimensional family of 
prescriptions $p$ and associated Hamiltonian constraints $\hat{H}^p(N)$
such that if 
$f_\gamma$ is a function cylindrical with respect 
to a graph $\gamma$ then there are projections
$\hat{H}^p(N)f_\gamma=\hat{H}^{p(\gamma)}_\gamma(N)f_\gamma
=\sum_{v\in V(\gamma)}N(v)\hat{H}^p_{v,\gamma}$ where 
$\hat{H}^{p(\gamma)}_{v,\gamma}$ is an operator which attaches certain 
edges in 
between pairs of edges of $\gamma$ incident at the vertex $v$ in a 
prescription dependent and graph dependent way. However, that 
prescription is covariantly defined, that is, if we map the graph with a 
diffeomorphism $\varphi$ then the prescription $p(\varphi(\gamma))$ is 
such that the operators $\hat{U}(\varphi)\hat{H}^{p(\gamma)}_{v,\gamma}$
and $\hat{H}^{p(\varphi(\gamma))}_{v,\varphi(\gamma)}$ are related by a 
diffeomorphism which in general differs from the identity. This has the 
important consequence that the dual of the
operators $\hat{H}^p(N)$, when evaluated on diffeomorphism invariant 
distributions, are independent of $p$, at least within 
the same diffeomorphism equivalence class of prescriptions $p$. Thus,
with this restriction, the dual operator $\hat{H}'(N)$ defined by
$\hat{H}'(N)\Psi(f):=\Psi(\hat{H}^p(N)f)$ is completely independent of
the choice of $p$, provided that $\Psi\in\Phi'$ is a diffeomorphism 
invariant distribution. Now, the problem
is that while $\Psi$ is diffeomorphism invariant, $\hat{H}'(N)\Psi$
is not. In fact we have the following theorem which demonstrates that
the commutator algebra between dual diffeomorphism and dual Hamiltonian 
constraint
closes in the way expected from the classical Poisson constraint algebra
$\{H(M),V(\vec{N})\}=H({\cal L}_{\vec{N}M})$ where $V$ is the classical
diffeomorphism constraint.
\begin{Theorem} \label{th3}
The quantum constraint algebra between the dual Hamiltonian constraint
operator $\hat{H}'(N)$ and the exponentiated diffeomorphism constraint
operator $\hat{U}(\varphi)$ is anomaly-free on diffeomorphism invariant
distributions $\Psi$, that is,
\be \label{8}
([\hat{H}(N),\hat{U}(\varphi)])'\Psi=\hat{H}'(\varphi^\star N-N)\Psi\;. \ee
\end{Theorem}
Proof :\\
Notice that it makes sense to check the constraint algebra only on 
diffeomorphism invariant elements of $\Phi'$ because this is where 
$\hat{H}'(N)$ is defined. Moreover, since $(\hat{U}(\varphi))'\Psi=\Psi$
it makes sense to compute the commutator.\\
So let $f_\gamma$ be a function in $\Phi$ cylindrical with respect to a graph
$\gamma$. Then we have by definition
\ba \label{9}
&&-(([\hat{H}(N),\hat{U}(\varphi)])'\Psi)(f_\gamma)
=([\hat{H}'(N),\hat{U}'(\varphi)]\Psi)(f_\gamma)\nonumber\\
&=&([\hat{H}'(N)-\hat{U}'(\varphi)\hat{H}'(N)]\Psi)(f_\gamma)
=\Psi([\hat{H}^p(N)-\hat{H}^p(N)\hat{U}(\varphi)]f_\gamma)
\nonumber\\
&=&\Psi(\hat{H}^p(N)f_\gamma-\hat{H}^p(N)f_{\varphi(\gamma)})
\nonumber\\
&=&\sum_{v\in V(\gamma)} 
\{N(v)\Psi(\hat{H}^{p(\gamma)}_{v,\gamma}f_\gamma)  -N(\varphi(v))
\Psi(\hat{H}^{p(\varphi(\gamma))}_{v,\varphi(\gamma)}
f_{\varphi(\gamma)}) \} \nonumber\\
&=&\sum_{v\in V(\gamma)} [N(v)-N(\varphi(v))]
\Psi(\hat{H}^{p(\gamma)}_{v,\gamma}f_\gamma) \nonumber\\ 
&=&(\hat{H}'(N-\varphi^\star N)\Psi)(f_\gamma)\;.
\ea
The only non-trivial step in this computation has been the last one in which
we have used the fact mentioned before that the 
prescription is covariantly defined and that the images of the 
Hamiltonian constraint on diffeomorphic vectors are diffeomorphic so 
that the difference is not seen by $\Psi$.\\
$\Box$\\
We also recall the following theorem from \cite{almmt}.
\begin{Theorem} \label{th4}
The commutator algebra of the exponentiated diffeomorphoism constraints
closes, that is,
\be \label{10a}
\hat{U}(\varphi)\hat{U}(\varphi')\hat{U}^{-1}(\varphi)\hat{U}^{-1}(\varphi')
=\hat{U}(\varphi\circ\varphi'\circ\varphi^{-1}\circ(\varphi)')^{-1})\;.
\ee
\end{Theorem}
Proof :\\
The proof follows trivially from the fact that $\hat{U}(\varphi)f_\gamma
=f_{\varphi(\gamma)}$ which in particular shows that $\hat{U}$ is a 
unitary representation of the diffeomorphism group on $\cal H$.\\
$\Box$\\
Theorems (\ref{th3}),(\ref{th4}) mean that the quantum Dirac algebra 
closes in 
the appropriate sense between Diffeomorphism constraints among 
themselves on 
${\cal H}$ and between Diffeomorphism constraints and the dual of the 
Hamiltonian constraint on ${\cal H}_{Diff}$.\\
What remains to be shown is that the algebra closes between two (duals of)
Hamiltonian constraints. In fact, the following theorem was proved in
\cite{1b}.
\begin{Theorem} \label{th5}
The quantum constraint algebra of the Hamiltonian constraint closes, 
that is,
\be \label{10}
(([\hat{H}(M),\hat{H}(N)])'\Psi)(f_\gamma)=0
\ee
for any diffeomorphism invariant $\Psi\in \Phi'$, any $f_\gamma\in \Phi$
and any $M,N$.
\end{Theorem}
Formula (\ref{10}) is unsatisfactory for two reasons : \\
1) On the left hand side one would like to see something like
$[\hat{H}'(M),\hat{H}'(N)]$ rather than
$([\hat{H}(M),\hat{H}(N)])'$.\\
2) On the right hand side one would like to see something like
$(\int d^3x (M_{,a} N-M N_{,a})\hat{q}^{ab}\hat{V}_b)'$ rather than just 
zero.\\ 
The problem with issue 1) is that the product operator 
$\hat{H}'(M)\hat{H}'(N)$ does not exist on ${\cal H}_{Diff}$. The 
problem with issue 2) is that none of the operators $\hat{q}^{ab},\hat{V}_b$
exist even on $\cal H$. Moreover, even if the right hand side would be 
of that kind, how would one tell the difference between this operator and
the zero operator ? The point is that whenever the algebra is anomaly free
then the right hand side must annihilate a diffeomorphism invariant state.\\
In the sequel we will somewhat improve (\ref{th4}) as follows : \\
First we emphasize that the prescription $p$ of \cite{1b} was {\em designed} 
such that not only $\hat{H}'(N)\Psi$ but
also the dual of arbitrary products $\hat{H}^p(N_1)..\hat{H}^p(N_n)$,
evaluated on diffeomorphism invariant states,
for arbitrary $N_i$ is independent of $p$, that is, there exists
a well-defined operator $[\hat{H}(N_1)..\hat{H}(N_n)]'$. This follows 
immediately from the fact that in \cite{1b} the prescription was
covariantly defined, that is, the prescription applied to graphs that
differ by a diffeomorphism $\varphi$ yields graphs that differ by 
{\em some} diffeomorphism $\varphi'$ which does not need to equal 
$\varphi$. We stress this point here one more
time for the sake of clarifying things, however, it is already contained in
\cite{1b}. This fact can be used 
to {\em define} products of dual operators {\em indirectly} to be the 
dual of the 
product and this was actually done in Theorem 3.1 and section 3.4
of \cite{1b}.
(For a different approach that involves a certain enlargement of the
space of diffeomorphism invariant distributions and therefore enables to 
define products of dual operators {\em directly} see \cite{LM}). 
Thus, we are able to remedy issue 1) above. As 
expected, the dual operator algebra turns out to be completely Abelian as 
one can immediately see from Theorem \ref{th5} and as pointed out in 
\cite{1a}.\\
Secondly, it turns out that one can quantize the classical expression
$\int d^3x (M_{,a} N-M N_{,a})q^{ab}V_b$ in the sense that one can turn it 
into a densely defined operator $\hat{O}(M,N)$ on $\cal H$. We then show 
that $\hat{O}'(M,N)\Psi=0$ for any diffeomorphism invariant $\Psi\in\Phi'$.
This remedies issue 2) above.\\
In conclusion we can write the operator equality (on ${\cal H}_{Diff}$)
$[\hat{H}'(M),\hat{H}'(N)]=\hat{O}(M,N)$ which is as close as we can get
to representing faithfully the quantum Dirac algebra in the present 
framework which suffers from the unavoidable fact that $\hat{H}(N)$ is 
state dependently defined (in other words, $A_a^i$ is not defined, only 
holonomies along paths are defined and the choice of path is the source of
the ambiguity). 

We begin with issue 1).
\begin{Theorem} \label{th6}
Let $\Psi\in\Phi'$ be a diffeomorphism invariant state and $f\in\Phi$.
Then $\Psi(\hat{H}^p(N_1)..\hat{H}^p(N_n)f)$ depends only on the 
diffeomorphism invariant properties of  the prescription $p$ for any
$N_1,..,N_n$ and any $n=1,2,..$.
\end{Theorem}
Proof :\\
We can assume that $f$ is cylindrical with respect to a graph $\gamma$.
Also, the ambiguity is already completely encoded in the precription 
dependence of the Euclidean Hamiltonian constraint $\hat{H}^p_E$ because the 
Lorentzian constraint is just a polynomial of these with operator
valued coefficients that leave the graph invariant \cite{1b}.\\
Now, whenever $\hat{H}^p_E$ acts on a function cylindrical with respect 
to a graph $\gamma$ then it reduces to a sum of terms 
$\hat{H}^p_{E,v,e,e'}$,
one for each vertex $v\in V(\gamma)$ and one for each pair $e,e'\in 
E(\gamma)$ of edges of $\gamma$ incident at $v$.  
The prescription $p$ assigns to each $\hat{H}^p_{E,v,e,e'}$,
{\em proper} segments $s\subset e,s'\subset e'$ incident at $v$ and an arc 
$a$ incident
at the endpoints of $s,s'$ and not intersecting the graph anywhere else 
together with a diffeomorphism invariant condition about the topology of 
the routing of $a$ through the edges of $\gamma$ incident at $v$.
The ``length" of $s,s'$ and the ``shape" of $a$ is the whole source of the 
precription dependence.
The point is that when $\hat{H}^p_E$ acts again, the new arcs to be 
attached {\em will never intersect the old arcs} already attached because
at each stage $s,s'$ are always proper segments of $e,e'$. Consider now
a particular term 
\be \label{exp}
\Psi(\hat{H}^p_{E,v_1,e_1,e_1'}\hat{H}^p_{E,v_2,e_2,e_2'}\;...\;
\hat{H}^p_{E,v_n,e_n,e_n'}f)
\ee
in the expansion of $\Psi(\hat{H}^p_E(N_1)..\hat{H}^p_E(N_n)f)$. If it 
does not vanish then $\Psi$ is a linear combination of group averaged 
cylindrical functions and one of them, $\Psi'$, depends on the graph 
$\gamma'=\tilde{\gamma}\cup a_1\cup..\cup a_n$ and it is the only term of 
$\Psi$ that contributes to (\ref{exp}). Here 
$\tilde{\gamma}\subset\gamma$ is the graph $\gamma$ with possibly some 
of the $s_i,s'_i$ removed \cite{1b,1c}. Now, since 
we allow for smooth diffeomorphisms that leave a graph analytic, there
are smooth diffeomorphisms involved in the definition of $\Psi'$ that leave
$\gamma$ invariant and $\gamma'$ analytic and that alter the 
length of $s_i,s_i'$ and the shape of $a_i$ arbitrarily {\em while keeping
the topology of $\gamma'$ invariant} (arcs cannot intersect each other
or $\gamma$ other than in their endpoints but they can slide up and down 
along $e,e'$ and change their shapes).
It follows that (\ref{exp}) does not depend on $p$ but only on its 
diffeomorphism class $[p]$.\\
$\Box$
\begin{Corollary} \label{col1}
The quantitity $[\hat{H}(N_1)..\hat{H}(N_n)]'\Psi$ defined by
$\Psi(\hat{H}^p(N_1)..\hat{H}^p(N_n)f)$ is well-defined, 
that is, only depends on the (fixed) diffeomorphism class $[p]$.
\end{Corollary}
\begin{Definition}
Let $\cal B$ be the vector space of sums of products of operators 
$\hat{H}^p(N)$ for some $p$ and let ${\cal B}'$ be the set of dual operators
defined by ${\cal B}'=\{B';\;\exists B\in {\cal B}\ni [B'\Psi](f)=\Psi(Bf)
\forall \Psi\in\Phi'_{Diff},\;f\in\Phi\}$ where again $\Phi'_{Diff}$ is the 
diffeomorphism invariant subset of $\Phi'$. We turn the vector space
${\cal B}'$ into an associative algebra by defining a product by
$B_1' B_2':=(B_2 B_1)'$. Notice that the product is a well-defined map from
${\cal B}'\times{\cal B}'$ to ${\cal B}'$ by Corollary (\ref{col1}).
\end{Definition}
\begin{Corollary} \label{col2}
The algebra ${\cal B}'$ is Abelian.
\end{Corollary}
Proof :\\
Notice that $[B_1',B_2']=([B_2,B_1])'$ and write $\Psi([B_2,B_1]f)$ as sums 
of expressions of the form $\Psi(B[H^p(M),\hat{H}^p(N)]f')$ where 
$B\in {\cal B}$ and $f'\in \Phi$. Now use Theorem (\ref{th6}) in the 
argument of the proof of Theorem (\ref{th5}) (or 3.1 in \cite{1b}) in 
order to complete the argument. (Essentially, one needs to show that one can 
slide the arcs along $e,e'$ despite the presence of the operator $B$.
This can be accomplished by straightforward application of Theorem 
(\ref{th6})).\\
$\Box$\\
\\
Next we turn to issue 2) :
In the last part of this section we show that one can indeed define an 
operator corresponding to the right hand side of $\{H(M),H(N)\}$ whose dual
annihilates diffeomorphism invariant distributions.

Let $\omega_a:=M N_{,a}-M_{,a} N$ and consider the triple of vector fields 
$v_a$ with components $(v_a)^b:=\delta_a^b$ as well as the triple of 
one forms $\sigma^b$ with components $(\sigma^b)_a:=\delta_a^b$.
Let $\chi_\epsilon(x,y)$ be the characteristic function of a coordinate box
with center $x$ and coordinate volume $\epsilon^3$. The volume of the box
as measured by $q_{ab}$ is given by $V(x,\epsilon):=\int d^3y 
\chi_\epsilon(x,y)\sqrt{\det(q)}(y)$. Finally, let 
$V_a=\mbox{tr}(F_{ab} E^b)$ be the classical infinitesimal generator of
diffeomorphisms.\\
With this preparation we have the following classical identity
\ba \label{17}
O(M,N)&:=&\int d^3x (\omega_a q^{ab}V_b)(x)\nonumber\\
&=&\lim_{\epsilon\to 0}\frac{1}{4\epsilon^6}
\int d^3x (\omega_a\epsilon^{abc}\epsilon_{ijk}
\frac{e_b^j e_c^k}{\sqrt{\det(q)}})(x)\times\nonumber\\
&& \times \int d^3y ((\sigma^g)_d\epsilon^{def}\epsilon_{ilm}
\frac{e_e^l e_f^m}{\sqrt{\det(q)}})(y)
\int d^3z \chi_\epsilon(x,z)\chi_\epsilon(y,z)
((v_d)^h V_h)(z) \nonumber\\             
&=&\lim_{\epsilon\to 0}\frac{\epsilon_{ijk}\epsilon_{ilm}}{4\epsilon^6}
\int  (\omega\wedge\frac{e^j\wedge e^k}{\sqrt{\det(q)}})(x)\times\nonumber\\
&& \times \int  ((\sigma^a)\wedge\frac{e^l\wedge e^m}{\sqrt{\det(q)}})(y)
\int d^3z \chi_\epsilon(x,z)\chi_\epsilon(y,z)((v_a)^b V_b)(z) \nonumber\\
&=&\lim_{\epsilon\to 0}
\frac{\epsilon_{ijk}\epsilon_{ilm}}{16\kappa^2\epsilon^6}
\int (\omega\wedge\frac{\{A^j,V(x,\epsilon)\}\wedge 
\{A^k,V(x,\epsilon)\}}{\sqrt{\det(q)}})(x)\times\nonumber\\
&& \times \int ((\sigma^a)\wedge
\frac{\{A^l,V(y,\epsilon)\}\wedge \{A^m,V(y,\epsilon)\}}{\sqrt{\det(q)}})(y)
\times\nonumber\\
&&\times \int d^3z \chi_\epsilon(x,z)\chi_\epsilon(y,z)((v_a)^b V_b)(z) 
\nonumber\\
&=&\lim_{\epsilon\to 0}
\frac{\epsilon_{ijk}\epsilon_{ilm}}{16\kappa^2}
\int (\omega\wedge\frac{\{A^j,V(x,\epsilon)\}\wedge 
\{A^k,V(x,\epsilon)\}}{V(x,\epsilon)})(x)\times\nonumber\\
&& \times \int ((\sigma^a)\wedge
\frac{\{A^l,V(y,\epsilon)\}\wedge \{A^m,V(y,\epsilon)\}}{V(y,\epsilon)})(y)
\times\nonumber\\
&&\times\int d^3z \chi_\epsilon(x,z)\chi_\epsilon(y,z)((v_a)^b V_b)(z) 
\nonumber\\ 
&=&\lim_{\epsilon\to 0}
\frac{\epsilon_{ijk}\epsilon_{ilm}}{\kappa^2}
\int (\omega\wedge\{A^j,\sqrt{V(x,\epsilon)}\}\wedge 
\{A^k,\sqrt{V(x,\epsilon)}\})(x)\times\nonumber\\
&& \times \int ((\sigma^a)\wedge
\{A^l,\sqrt{V(y,\epsilon)}\}\wedge 
\{A^m,\sqrt{V(y,\epsilon)}\})(y)\times\nonumber\\
&&\times\int d^3z \chi_\epsilon(x,z)\chi_\epsilon(y,z)((v_a)^b V_b)(z) 
\ea
where in the first step we just have written $q^{ab}$ in terms of $e_a^i$,
in the second step we have rewritten everything interms of forms, in the 
third step we have used the key identity \cite{1a,1b}
$$
2\kappa\mbox{sgn}(\det((e_b^j)))e_a^i=\{A_a^i(x),V(x,\epsilon)\}\;,
$$
in the fourth step we used that $V(x,\epsilon)/\epsilon^3\to 
\sqrt{\det(q)}(x)$
as $\epsilon\to 0$ and finally in the last step we just used 
$\{A_a(x),\sqrt{V(x,\epsilon)}\}=\{A_a(x),V(x,\epsilon)\}/
(2 \sqrt{V(x,\epsilon)})$. 

Expression (\ref{17}) is just one example of a 
general theorem proved in \cite{2} which says that singular prefactors 
of $\sqrt{\det(q)}$ can always be absorbed into Poisson brackets which
then are subject to quantization as follows :\\
Given a function $f_\gamma$ cylindrical with respect to a graph $\gamma$
we triangulate $\Sigma$ in adaption to $\gamma$ as in \cite{1b}. We leave the
the quantization $\hat{V}_a$ of $V_a$ open at this point and will specify 
it only later. For the moment being we just order it to the left hand side
and replace Poisson brackets by commutators times $1/(i\hbar)$. The 
connection $A$ gets replaced, just as in \cite{1b}, by a holonomy along 
the edges of a tetrahedron. We use a basis of $su(2)$ with the 
algebra $[\tau_i,\tau_j]=\epsilon_{ijk}\tau_k$ so that $\mbox{tr}(\tau_i A)
=-A^i/2$. Each expression of the form 
$\mbox{tr}(\tau_i h_s[h_s^{-1},\hat{V}^{1/2}(v,\epsilon)])f_\gamma$, where 
$v$ is a vertex of $\gamma$ and $s$ a segment of an edge starting at $v$,
does not depend on $s$ because it must be gauge invariant at the endpoint 
of $s$. Since the volume operator only acts at vertices of $\gamma$
we find in complete analogy with computations in \cite{1b} the regulated
operator
\ba \label{18}
\hat{O}_\epsilon(M,N)f_\gamma&=&
\frac{16\epsilon_{ijk}\epsilon_{ilm}}{\hbar^4\kappa^2}\sum_{v,v'\in 
V(\gamma)}
\int d^3z \chi_\epsilon(v,z)\chi_\epsilon(v',z)((v_a)^b \hat{V}_b)(z) 
\times\nonumber\\
& \times& \sum_{v(\Delta)=v}\epsilon^{npq}
[M(v)N(s_n(\Delta))-M(s_n(\Delta))N(v)]\times\nonumber\\
&\times& \mbox{tr}(\tau_j h_{s_p(\Delta)}
[h_{s_p(\Delta)}^{-1},\hat{V}^{1/2}(v,\epsilon)])
\mbox{tr}(\tau_k h_{s_q(\Delta)}
[h_{s_q(\Delta)}^{-1},\hat{V}^{1/2}(v,\epsilon)])\times\nonumber\\
& \times& \sum_{v(\Delta')=v'}\epsilon^{rst}
(\sigma^a)_c[s_r^c(\Delta')-v^{\prime c}]\times\nonumber\\
&\times&
\mbox{tr}(\tau_l h_{s_s(\Delta')}
[h_{s_s(\Delta')}^{-1},\hat{V}^{1/2}(v',\epsilon)])
\mbox{tr}(\tau_m h_{s_t(\Delta')}
[h_{s_t(\Delta')}^{-1},\hat{V}^{1/2}(v',\epsilon)])f_\gamma\nonumber\\
&&
\ea
where in abuse of notation $s_n(\Delta)$ denotes also the endpoint of the 
segment $s_n(\Delta)$ whose starting point is $v(\Delta)$.\\
Fix the value of $a,v,v'$ and let now $\xi^b(z):=(v_a)^b
\chi_\epsilon(v,z)\chi_\epsilon(v',z)$. We replace $E^a_i$ by
$\hat{E}^a_i(z)=-i\hbar\kappa\delta/\delta A_a^i$ in $V_b$ and order 
$\hat{E}$ to the right hand side. Let $f'_\gamma$ be any function 
cylindrical with respect to $\gamma$. Also we abbreviate
$$
X^i_e(t):=[h_e(0,t)\tau_i h_e(t,1)]_{AB}\partial/\partial(h_e(0,1))_{AB} 
$$
for each edge $e$ of $\gamma$. Then we have 
\ba \label{19}
&& \int d^3x \xi^a(x)\hat{V}_a(x)f'_\gamma \nonumber\\
&=& -i\hbar\kappa\int d^3x \xi^a(x) F_{ab}^i(x)
\sum_{e\in E(\gamma)}\frac{\delta (h_e(0,1))_{AB}}{\delta A_b^i(x)}
\frac{\partial f'_\gamma}{\partial (h_e(0,1))_{AB}}
\nonumber\\
&=& -i\hbar\kappa\sum_{e\in E(\gamma)}\int d^3x \xi^a(x) F_{ab}^i(x)
\int_0^1 dt \delta(x,e(t)) \dot{e}^b(t) X^i_e(t)f'_\gamma
\nonumber\\
&=& -i\hbar\kappa\sum_{e\in E(\gamma)}\int_0^1 dt  \xi^a(e(t)) F_{ab}^i(e(t))
\dot{e}^b(t) X^i_e(t)f'_\gamma
\nonumber\\
&=& -i\hbar\kappa\sum_{e\in E(\gamma)}\lim_{n\to\infty}
\sum_{k=1}^n \xi^a(e(t_k)) F_{ab}^i(e(t_k))[e^b(t_k)-e^b(t_{k-1})] 
X^i_e(t_k)f'_\gamma 
\ea
for any partition $0=t_0<t_1<..<t_n=1$ of $[0,1]$ which we may choose 
individually for each edge $e$.\\
We now combine (\ref{18}) and (\ref{19}). We choose to take first the 
$\epsilon\to 0$ limit. (This exchange of limits is justified by the fact
that one can alternatively also keep $\epsilon\approx 1/n$ finite and 
and still arrive at the quantum result below, the virtue being that  
this way of guiding the limit also gives back the original function 
$O(M,N)$ on the classical side. See \cite{2,3} for details) . 
The product of charcteristic functions 
$\chi_\epsilon(v,e(t_k))\chi_\epsilon(v',e(t_k))$ will vanish for any $k,n$
if $v\not=v'$ when choosing $\epsilon$ sufficiently small. We conclude 
that only the terms with $v=v'$ survive in (\ref{18}). Next, since 
$v$ is a vertex for sufficiently small $\epsilon$, $\chi_\epsilon(v,e(t_k))$
vanishes (keeping $n$ fixed) unless $k=0$ because we have labelled edges 
$e$ such that only their starting point can be a vertex of $\gamma$. 
Therefore in the limit $\epsilon\to 0$ the sum over $k$ in (\ref{19}) 
disappears and the dependence on $\epsilon$ disappears when we replace 
$\hat{V}(v,\epsilon)$ by the operator $\hat{V}_v$ as explained in 
\cite{3}. Finally we contract the indices $a$ in $\xi^b(e(0))=\delta_a^b$
from (\ref{19}) and in $(\sigma^a)_c[s_r^c(\Delta')-v^c]=
[s_r^a(\Delta')-v^A]$ and find that (\ref{18}) becomes
\ba \label{20}
&& \hat{O}_n(M,N)f_\gamma= 
-i\frac{16\epsilon_{ijk}\epsilon_{ilm}}{\hbar^3\kappa}
\epsilon^{rst}\epsilon^{npq} \sum_{v\in V(\gamma)}\;\;
\sum_{e\in E(\gamma),e(0)=v}\;\; 
\sum_{v(\Delta)=v(\Delta')=v}\times\nonumber\\
&\times&[s_r^a(\Delta')-v^a] F_{ab}^i(v)[e^b(t_1)-v^b] X^i_e
[M(v)N(s_n(\Delta))-M(s_n(\Delta))N(v)]\times \nonumber\\
&\times& \mbox{tr}(\tau_j h_{s_p(\Delta)}
[h_{s_p(\Delta)}^{-1},\hat{V}^{1/2}_v])
\mbox{tr}(\tau_k h_{s_q(\Delta)}
[h_{s_q(\Delta)}^{-1},\hat{V}^{1/2}_v])\times\nonumber\\
& \times& 
\mbox{tr}(\tau_l h_{s_s(\Delta')}
[h_{s_s(\Delta')}^{-1},\hat{V}^{1/2}_{v'}])
\mbox{tr}(\tau_m h_{s_t(\Delta')}
[h_{s_t(\Delta')}^{-1},\hat{V}^{1/2}_v])f_\gamma
\ea
where we have used that the operators in the second and third line do not 
change the graph $\gamma$ and $X^i_e=X^i_e(0)$ is the right invariant 
vector field on $SU(2)$ evaluated at $h_e(0,1)$. Finally, let us choose 
$t_1$ for each $e$ 
individually such that $e(t_1)$ coincides with the segment of $e$ that 
appears as an edge $s_i(\Delta)$ of one or  
several tetrahedrons $\Delta$ of the triangulation with $v(\Delta)=v$
and let us call this segment $e(t_1):=s_i(\Delta)=:s(e)$. Then 
(\ref{20}) only depends on the triangulation $T$. The final step
is to replace the curvature expression 
$[s_r^a(\Delta')-v^a] F_{ab}(v)[s(e)^b-v^b]$ by 
$h_{\alpha(s_r(\Delta'),s(e))} -1_{SU(2)}$ where 
$\alpha(s_r(\Delta'),s(e))$ is the loop $s_1\circ s_2\circ s_3^{-1}
\circ s_4^{-1}$ where $s_1=s_r(\Delta'), s_4=s(e)$ and $s_2,s_3$ 
respectively are some arbitrarily chosen segments which do not intersect 
$\gamma$ except in the end points of $s_1,s_4$ such that the loop 
$\alpha$ looks like a little parallelogramm in the coordinate frame under
consideration. Moreover, the ``arc with a corner" $s_2\circ s_3^{-1}$ is 
subject to the same routing conditions as the arcs of tetrahedra (see 
\cite{1b} for details).

Now consider the combination
$$
[s_r^a(\Delta')-v^a] F^i_{ab}(v)[s(e)^b-v^b] X^i_e\approx
\mbox{tr}(h_{\alpha(s_r(\Delta'),s(e))}h_e(0,1)\partial_{h_e(0,1)})
-\mbox{tr}(h_e(0,1)\partial_{h_e(0,1)})
$$
and compare it with the infinitesimal variation of $f'_\gamma$ under the 
deformation of $e=s(e)\circ e'$ into 
$\tilde{e}=\alpha(s_r(\Delta'),s(e))\circ e=s_1\circ s_2\circ 
s_3^{-1}\circ e'$. We find (we suppress all arguments different from $h_e$)
\ba \label{21}
\delta f'_\gamma&:=&f'_\gamma(h_{\tilde{e}})-f'_\gamma(h_e)
=f'_\gamma(h_{\alpha(s_r(\Delta'),s(e))}h_e)-f'_\gamma(h_e)\nonumber\\
&=& f'_\gamma(h_e+[h_{\alpha(s_r(\Delta'),s(e))}-1_{SU(2)}]h_e)
-f'_\gamma(h_e)\nonumber\\
&=& \mbox{tr}([h_{\alpha(s_r(\Delta'),s(e))}-1_{SU(2)}]h_e\partial_{h_e})
f'_\gamma +o([h_{\alpha(s_r(\Delta'),s(e))}-1_{SU(2)}]^2)\nonumber\\
&=&[\hat{U}(\varphi(s_r(\Delta'),s(e)))-\mbox{id}_{\cal H}]f_\gamma
+o([h_{\alpha(s_r(\Delta'),s(e))}-1_{SU(2)}]^2)
\ea
where in the last step we have written the variation of $f_\gamma$ as 
the change under a corresponding diffeomorphism $\varphi(s_r(\Delta'),s(e))$.
Clearly, such a diffeomorphism will not only deform $e$ but also the edge
of $\gamma$ which has $s_r(\Delta')$ as a segment. However, one may 
choose $\varphi(s_r(\Delta'),s(e))$ such that the corresponding change of 
$f'_\gamma$ is of higher order.

Expression (\ref{21}) shows that (up to higher order terms which vanish 
in the continuum limit) the operator $\hat{V}_b$ can be 
multiplicatively regulated by multiplying it with a small but finite 
line element (here given by $\xi^a$) and writing it as 
a small but {\em finite} diffeomorphism minus the identity operator.
This is because the infinitesimal operator does not make any sense on $\cal 
H$ as mentioned earlier and since $\xi^a$ comes from $q^{ab}$, it happened
what was to be expected : While $q^{ab},V_b$ individually do not make
sense as operators, the combination $q^{ab} V_b$ in fact does make sense.\\
Thus, combining (\ref{20}) and (\ref{21}) we obtain the desired end result
($T$ reminds us of the triangulation dependence)
\ba \label{22}
\hat{O}_T(M,N)f_\gamma &=& 
-i\frac{16\epsilon_{ijk}\epsilon_{ilm}}{\hbar^3\kappa}
\epsilon^{rst}\epsilon^{npq} \sum_{v\in V(\gamma)}
\sum_{e\in E(\gamma),e(0)=v} \sum_{v(\Delta)=v(\Delta')=v}\times\nonumber\\
&\times& [\hat{U}(\varphi(s_r(\Delta'),s(e)))-\mbox{id}_{\cal H}]
[M(v)N(s_n(\Delta))-M(s_n(\Delta))N(v)]\times\nonumber\\
&\times& \mbox{tr}(\tau_j h_{s_p(\Delta)}
[h_{s_p(\Delta)}^{-1},\hat{V}^{1/2}_v])
\mbox{tr}(\tau_k h_{s_q(\Delta)}
[h_{s_q(\Delta)}^{-1},\hat{V}^{1/2}_v])\times\nonumber\\
& \times& 
\mbox{tr}(\tau_l h_{s_s(\Delta')}
[h_{s_s(\Delta')}^{-1},\hat{V}^{1/2}_{v'}])
\mbox{tr}(\tau_m h_{s_t(\Delta')}
[h_{s_t(\Delta')}^{-1},\hat{V}^{1/2}_v])f_\gamma\;.
\ea
The only point which is important is that that the diffeomorphism operator\\
$\hat{U}(\varphi(s_r(\Delta'),s(e)))$ stands {\em to the left} in (\ref{22})
which implies that when evaluating (\ref{22}) on a diffeomorphism invariant
distribution $\Psi$, the result vanishes. That such an ordering is 
possible while still obtaining a (precription dependent) densely defined 
operator is nevertheless a non-trivial result.\\
Moreover, $\hat{O}_T(M,N)$ is 
clearly just like $\hat{H}(M)$ prescription dependent but its dual 
$\hat{O}'(M,N)$ is not. We may therefore simply set 
$[\hat{H}'(N),\hat{H}'(M)]=\hat{O}'(M,N)$ on $\Phi'_{Diff}$.

Formula (\ref{22}) deserves several comments : 
\begin{itemize}
\item[1)] The immediate question is : Can we do even better than this ?
That is, can one recover (\ref{22}) by 
simply evaluating the commutator $[\hat{H}(M),\hat{H}(N)]f_\gamma$ ?
Comparing (\ref{22}) with the expression found in \cite{1b} for the 
commutator shows some similarities but also some key differences :\\
The commutator, at least for the Euclidean Hamiltonain constraint, involves 
two factors of operators $\hat{V}$ while (\ref{22}) involves four factors of
$\hat{V}^{1/2}$. Thus, while the overall power is the same, the factor 
ordering is different. Next, the commutators involve two loops but 
(\ref{22}) only one. Checking on generic states $f_\gamma$ reveals that 
the two loops do not in general combine into only one loop. Trial and error
reveals that one has a chance to combine the two loops into one 
if and only if the Hamiltonian constraint acts also at the vertices it 
creates. However, then one finds that when evaluating on diffeomorphism 
invariant states, the result does not vanish any longer. 

This is therefore a major difference which seems insurmountable and 
shows that the commutator and (\ref{22}) are simply
very different quantizations of the same classical object (meaning that,
since the classical expression $O(M,N)$ can be written as a Poisson 
bracket, we may choose to quantize it through the corresponding 
commutator). The deeper reason for this difference is the following : 
when one does the classical 
computation to prove that indeed $\{H(M),H(N)\}=O(M,N)$,  
one has to do several rather intricate algebraic
manipulations and integrations by part. In particular, the reason why one 
curvature term disappears is because the Poisson bracket picks the 
partial derivative of $A$ in $F$. None of these manipulations can have a 
quantum analogue because $F,A$ are not well-defined operators on $\cal H$.
It is therefore not surprising that the commutator does not give (\ref{22}).
Finally, it is reassuring that the combination 
$[M(v)N(s_n(\Delta))-M(s_n(\Delta))N(v)]$ in 
(\ref{22}) {\em does} appear in the commutator.
\item[2)] That $\hat{O}(M,N)$ can be given a meaning at all is astonishing at
first because we know that the infinitesimal generator of $\hat{U}(\varphi)$
as well as $\hat{q}^{ab}$ do not exist on $\cal H$. However, since both
operators combine to a density of weight {\em one}, the structure theorem 
proved in \cite{2} applies and it is therefore clear that the {\em 
integrated} expression $O(M,N)$ can be promoted to a densely defined 
operator on $\cal H$.
\item[3)] Concluding, we see that the quantum Dirac algebra closes in the 
appropriate sense and gives a faithful representation of the classical
Dirac algebra in the maximal possible way given the choice of representation 
$\cal H$.
\end{itemize}

\section{$\Theta$ moduli, Superselection Sectors and Separability}

In this short section we mainly wish to express a speculation which
needs further investigation to make it definite. The issue is the 
following :\\
Notice that we have group averaged the states with respect to (subsets of)
smooth diffeomorphisms and that graphs were piecewise analytic. Consider a 
graph with an $n-$valent vertex $v$.
The diffeomorphism group at $v$ reduces to a $GL(3,\Rl)$ matrix and so 
one has at most $d^2=9$ free parameters to change the tangent directions 
of all $n$ edges incident at $v$. The number of angles $\theta$ (with 
respect to some background metric) between those tangents is given by 
$n(n-1)/2$ which for $n\ge 5$ exceeds $9$ : This is the first example of
the appearance of the
$\theta$ moduli, that is, there is diffeomorphism invariant information
encoded in the intersections of analytic edges. 
These moduli take values in a compact subset of $\Rl$ with the cardinality
of $\Rl$. It is these $\theta$ moduli which make ${\cal H}_{Diff}$ a 
non-separable Hilbert space ! Other kinds of moduli which 
are invariant under smooth, that is, $C^\infty$ diffeomorphisms are the
the degrees of differentiability of intersections of analytic edges, however,
these moduli are integer valued and therefore do no spoil separability.

The question arises whether these moduli have any physical significance 
or are just an artefact of the fact that we are dealing with 
piecewise analytic graphs and smooth analyticity preserving 
diffeomorphisms. In fact, in \cite{Z} it is shown that if one works, for 
instance, with piecewise linear graphs or simply enlarges the 
diffeomorphism group, then the moduli disappear. Another striking feature 
of the formalism is that geometric operators like length, area and volume 
which one would expect to depend on those moduli actually do not depend 
on them as pointed out in \cite{TTL} which is an immediate consequence 
of the fact that their spectrum is discrete. In fact, the only diffeomorphism
invariant moduli that play a role are orientations of certain tangent
directions and so are of the discrete type \cite{Geometry}.
The speculation is that none of the physical operators can detect the 
continuous moduli. By this we mean the following  :\\ 
Notice that the number of moduli parameters is countable : a 
concrete counting 
scheme is provided by numbering graphs by some integers corresponding
to the number of vertices together with their valences (this fixes the
number of edges) some integers that fix the routing of edges between
vertices and some integers $k$ that label $C^k$ properties of 
differentiable intersections of analytic edges (we do not need continuous 
parameters labelling the germs of analytic edges because we average with 
respect to analyticity preserving smooth diffeomorphisms rather than 
analytic diffeomorphisms). Let 
$\underline{\theta}:=(\theta_1,\theta_2,..)$ denote a sequence of such a 
complete set of moduli. Since functions cylindrical with respect to graphs 
labelled with different moduli parameters are orthogonal with respect 
to $\cal H$ they continue to be orthogonal with respect to ${\cal H}_{Diff}$
by the very definiton of the group averaging proposal. 
Thus, the Hilbert space ${\cal H}_{Diff}$ breaks up into a continuous direct
sum
\be \label{dirsum}
{\cal H}_{Diff}=\oplus_{\underline{\theta}} 
{\cal H}_{Diff}^{\underline{\theta}}
\ee
and each of the Hilbert spaces 
${\cal H}_{Diff}^{\underline{\theta}}$ corresponds to a fixed sequence
$\underline{\theta}$, all of them being naturally isomorphic to each other.
Notice that the Hamiltonian constraint does not mix these {\em sectors}
because it attaches arcs to the graph in such a way that the new 
vertices are tri-valent for which there are no continuous moduli.
Moreover, ${\cal H}_{Diff}^{\underline{\theta}}$ is separable : we have
indicated above how to count graphs and otherwise the set of labellings of 
edges and vertices with spins and contractors also is countable. Thus,
the corresponding spin-network states provide an explicit orthonormal 
and countable basis.\\
\\
{\it $\Theta$-Superselection Assumption\\
The continuous moduli are superselected, that is, no physical observable
maps between states of different sectors
${\cal H}_{Diff}^{\underline{\theta}}$.}\\
\\
Summarizing, if the $\Theta$-Superselction Assumption turns out to be correct
(and the present results indicate that this might be the case) then 
we may restrict ourselves to a {\em separable} Hilbert space by fixing
a value of $\underline{\theta}:=\underline{\theta}_0$. This means, for 
instance, that if an 
operator creates new vertices on a graph with valence five or higher then
it must create them with the correct value of the moduli corresponding 
to our choice $\underline{\theta}_0$.

\section{An inner product on the space of physical solutions}

In \cite{1c} a general algorithm was displayed to construct {\em all} 
solutions
of the Lorentzian Hamiltonian constraint (and diffeomorphism constraint) of 
pure quantum general relativity. A huge number of solutions have the 
surprising feature that they are normalizable with respect to the inner
product for solutions of the diffeomorphism constraint. The treatment
was incomplete in the sense that no inner product for the solutions to 
all constraints was constructed which is capable to render normalizable also 
solutions which are not normalizable with respect to the inner product to
the diffeomorphism constraint. The reason for why such an inner product 
is not entirely straightforward to construct is because the favoured version
of the Hamiltonian constraint operator constructed in \cite{1b} is not 
self-adjoint
(also a symmetric version of the Hamiltonian constraint operator was 
constructed in \cite{1c} but it involves a rather unnatural technique
which makes it less appealing). This fact makes the group averaging
technique to construct an inner product as given in \cite{almmt,4,5}
inapplicable.

On the other hand, arguments first raised in \cite{6} suggest that 
$\hat{H}(M)$ {\em must not} be self-adjoint in order that the algebra closes.
In fact, the symmetric operator $\hat{H}'(M)+\hat{H}'(M)^\dagger$ is easily
seen to be anomalous.

We therefore conclude that we need to deal with the not self-adjoint 
operator $\hat{H}(M)$ and are asked to invent a group averaging technique
for not self-adjoint operators. The following quantum mechanical example 
will hint how that might be done in the case of interest.

\subsection{A two-oscillator example}

We consider the Hilbert space ${\cal H}:=L_2(\Rl,dx)\otimes L_2(\Rl,dy)$ 
corresponding to a system with two degrees of freedom. $\cal H$ plays the 
same role as $\Phi'_{Diff}$ does in quantum general relativity. In order 
to model the Hamiltonian constraint of quantum gravity it will be sufficient
for our purposes to consider only its Euclidean part $\hat{H}^E(M)$ which is 
simpler, at least to start with (we will generalize it later to the 
Lorentzian case). The 
most important property of $\hat{H}^E(M)$ is that it 
creates new edges when applying it to a function cylindrical with respect 
to a graph $\gamma$. Thus, we may think of $\hat{H}^E(M)$ as some kind of 
creation operator in harmonic oscillator terminology where the excitation
of the harmonic oscillator is replaced by an excitation of the 
gravitational field along the new edge. We therefore study the following
oversimplified model operator corresponding to $\hat{H}^E(M)$
\be \label{23}
\hat{C}:=\hat{a}^\dagger-\bar{c}\hat{b}^\dagger
\ee
where $c$ is an arbitrary complex number and $\hat{a},\hat{b}$ 
respectively are the usual annihilation operators associated with the 
$x,y$ coordinate respectively. Clearly the operator $\hat{C}$ is far from 
being self-adjoint on $\cal H$ and so we cannot use the usual group 
averaging procedure to construct an inner product on the space of its 
solutions. \\
However, this model is simple enough so that {\em all} the solutions to 
$\hat{C}$ can be constructed in {\em closed form}, just as in the 
gravitational case. The complete solution space will suggest a natural 
inner product theron. 

So, let $|m,n>$ be the usual harmonic 
oscillator eigenfunctions, that is, $\hat{a}|m,n>=\sqrt{m}|m-1,n>,\;
\hat{a}^\dagger|m,n>=\sqrt{m+1}|m+1,n>$ so that $[\hat{a},\hat{a}^\dagger]
=1$ and similarly for $\hat{b}$. In fact, $\hat{a}^\dagger$ almost precisely 
models the operator 
$\mbox{tr}([h_\alpha-h_\alpha^{-1}]h_s[h_s^{-1},\hat{V}])$
which is the basic building block of the Euclidean Hamiltonian constraint.
Here $\alpha$ is a a loop bounding a face of a tetrahedron 
$\Delta$ considered in previous sections and $s$ is an edge of $\Delta$.
Namely, $\hat{a}^\dagger$ does not only rise the excitation by one unit
(which corresponds to $h_\alpha^{\pm 1}$) but also multiplies by a 
coefficient $\sqrt{m+1}$ which depends on the ``spin" $m$ (corresponding
to $h_s[h_s^{-1},\hat{V}]$). Even the power $1/2$ of the spin is correct :
The volume operator $\hat{V}$ is bounded by 
$\sqrt{|\hat{J}^3|}$ where $\hat{J}$ is the total spin of the graph. 
Thus it should have eigenvlues of order $m^{3/2}$. Now the commutator 
acts roughly like a spin derivative, leaving us with a leading order of 
$m^{1/2}$.\\
The span of these oscillator 
eigenfunctions constitutes a topological vector space $\Phi$ which is 
dense in $\cal H$ and is the precise analogue of (diffeomorphism 
invariant) spin-network states with ``spins" $m,n$. We are interested in 
distributions $\Psi\in\Phi'$ (the dual of $\Phi$) where 
$\Phi\subset{\cal H}\subset\Phi'$
such that $\Psi(\hat{C}|m,n>)=(\hat{C}'\Psi)(|m,n>)=0$ for all $m,n$.
We are looking for $\Psi$'s of the form $\Psi(f):=\int dx dy 
\overline{\Psi}f$
and then the dual operator can be explicitly computed and is given by
$\hat{C}'=\hat{a}-c\hat{b}$. That is, it is a complex linear combination 
of annihilation operators, just as in the gravitational case. We can 
immediately see why $\hat{C}'$ has a huge number of solutions $\Psi$ 
which actually lie in $\Phi$ : Define the {\em level} of an oscillator 
state $|m,n>$ to be the number $m+n$. Then, if $f\in\Phi$ is a finite linear 
combination of oscillator states none of whose level exceeds a given 
number $N$, it follows that $(\hat{C}')^{N+1}f=0$ because $(\hat{C}')^{N+1}$
is a linear combination of monimials of the form $\hat{a}^{k}\hat{b}^{N+1-k}$
all of which reduce the level of an oscillator state by $N+1$ or 
annihilate it. In other words, $\hat{C}'$ is nilpotent on states of a
given level $N$.

Explicitly, the complete solution space can be constructed as follows : \\
The most general state of level $N$ is given by
\be \label{24}
|N,\{c\}>=\sum_{m=0}^N c_m |m,N-m>
\ee
and these states (over)span $\Phi$ as the $c_m$ take all complex values.
We apply $\hat{C}'$ and find
\ba \label{25}
&&\hat{C}'|N,\{c\}>
= \sum_{m=0}^N c_m [\sqrt{m} |m-1,N-m>-c\sqrt{N-m}|m,N-m-1>]\nonumber\\
&=& \sum_{m=1}^N c_m \sqrt{m} |m-1,N-m>
-c\sum_{m=0}^{N-1}c_m \sqrt{N-m}|m,N-m-1>]\nonumber\\
&=& \sum_{m=0}^{N-1} c_{m+1} \sqrt{m+1} |m,N-1-m>
-c\sum_{m=0}^{N-1}c_m \sqrt{N-m}|m,N-1-m>]\nonumber\\
&=& \sum_{m=0}^{N-1} [c_{m+1}\sqrt{m+1}-z c_m \sqrt{N-m}] |m,N-1-m>\;.
\ea
Since $\hat{C}'$ reduces the level precisely by one unit it follows that
the condition $(\hat{C}'\Psi)(|n,m>)=0$ is only a condition on those vectors
involved in the definition of $\Psi$ with the the same level. Thus 
any solution $\Psi$ is a (not necessarily finite) linear combination of 
vectors of the form (\ref{24}). We find the condition
\be \label{26}
c_{m+1}\sqrt{m+1}-c c_m \sqrt{N-m}=0\mbox{ or }
c_m=c^m\sqrt{\left( \begin{array}{c} N\\ m \end{array} \right) }c_0
\ee
and if we wish to normalize the solution then $|c_0|=(1+|c|^2)^{-N/2}$.
Let us simply denote the solution of level $N$ by $|N>$. 
Of course not every solution need to be normalizable since we check 
$(\hat{C}'\Psi)(f)$ only with elements $f\in\Phi$ so that there is no
growth restriction on the coefficients $d_N$ in $\Psi=\sum_{N=)}^\infty
d_N |N>$.

Another way of 
looking at what has happened is that we performed a Bogolubov transformation
on the annihilation operators and require that the physical vacuum is 
annihilated by one of the two new annhilation operators. The states $|N>$
are therefore nothing else than the excited states corresponding to the 
other (independent) annihilation operator.

We now address the question which inner product to impose on the solutions
$\Psi$ so obtained. 
Usually one expects that a typical solution to the Hamiltonian constraint
is not normalizable with respect to the inner product on $\cal H$. The 
intuition behind this expectation is experience with typical solutions
of self-adjoint quantum constraints that can be obtained by group averaging.
Let us give an example :\\ 
Consider the same Hilbert space as above but instead of the constraint
$\hat{C}$ consider the constraint $\hat{H}=\hat{p}-\hat{q}$ where 
$p,q$ are the momenta conjugate to $x,y$ respectively. Thus, $\hat{H}$ is 
a self-adjoint operator on $\cal H$ and we may follow the group average 
proposal to construct the space of solutions and an inner product thereon.
Denote by $|p,q>$ the usual momentum {\em generalized} eigenstates. 
Notice that these kinematical states, in contrast to the $|m,n>$, are already
not normalizable. Let now $|f>\in\Phi$. Then the group average map is given
by
\ba \label{27}
|\tilde{f}>&:=& \hat{\eta} |f>:=\int_\Rl 
\frac{dt}{2\pi}e^{it\hat{H}}|f>\nonumber\\
&=&\int_{\Rl^2} dp dq\int_\Rl \frac{dt}{2\pi}e^{it\hat{H}}|p,q><p,q|f>
=\int_{\Rl^2} dp dq \delta(p,q)|p,q><p,q|f>\nonumber\\
&=&\int_\Rl dp |p,p><p,p|f>,
\ea
that is, $\hat{\eta}=\int dp |p,p><p,p|$. Notice that $\hat{H}|p,p>=0$.\\
The group average inner product 
is defined by
\be \label{28}
<\tilde{f},\tilde{g}>_\sim:=<\tilde{f}|g>=<f,\hat{\eta} g>
=\int dp <f|p,p><p,p|g>\;.
\ee
In particular, we have $||\;|\tilde{f}>||_\sim^2=\int dp 
|<p,p|f>|^2<\infty$ since $||\;|f>||^2=\int dp dq |<p,q|f>|^2<\infty$
(under suitable conditions on $<p,q|f>$ which defines $\Phi$) but
$||\;|\tilde{f}>||^2=\int dp dq <p,p|f> <p,p|q,q><q,q|f>=\delta(0)
||\;|\tilde{f}>||_\sim^2=\infty$ blows up. This feature is shared by 
all models whose self-adjoint constraint operator has continuous 
and unbounded spectrum.\\
We will now study a second example with a self-adjoint, unbounded 
constraint 
operator whose spectrum is, however, discrete. Here we will encounter 
a surprise : all the group average solutions are normalizable elements of 
the Hilbert space.\\
Consider again the Hilbert space $\cal H$ as above but now the 
self-adjoint constraint operator is given by 
$\hat{H}:=\hat{a}^\dagger\hat{a}-\hat{b}^\dagger\hat{b}$, that is, the 
difference between two harmonic oscillator Hamiltonians. Clearly, the 
spectrum is entirely discrete and is given by the set of integers.
The oscillator eigenstates $|m,n>$ are eigenstates of $\hat{H}$ with 
eigenvalue $m-n$ and clearly any solution is a (not necessarily finite)
linear combination of the states $|m,m>$ for arbitrary $m\ge 0$. This is 
precisely what the group average map does as well : we have 
\ba \label{29}
|\tilde{f}>&:=&\hat{\eta}|f>:=\int_{S^1} \frac{dt}{2\pi}e^{it\hat{H}}|f>
\nonumber\\
&=&\sum_{m,n}\int_{-\pi}\pi \frac{dt}{2\pi}e^{it\hat{H}}|m,n><m,n|f>
=\sum_{m,n} \delta_{m,n} |m,n><m,n|f>
\nonumber\\
&=& \sum_m |m,m><m,m|f>
\ea
that is, $\hat{\eta}=\sum_m |m,m><m,m|$. The reason why $t$ now only 
ranges over $S^1$ is because the spectrum of $\hat{H}$ is discrete
in contrast to the above situation and so the group average must 
produce a Kronecker delta rather than a Dirac delta. Notice 
that the group averaging inner product is given by 
\be \label{30}
<\tilde{f},\tilde{g}>_\sim=\sum_m <f|m><m|g>
\ee
so that in particular\\ 
$||\;|\tilde{f}>||_\sim^2=\sum_m |<m,m|f>|^2
\le \sum_{m,n} |<m,n|f>|^2=||\;|f>||^2$ and \\
$||\;|\tilde{f}>||^2=\sum_{m,n} <m,m|f><m,m|n,n><n,n|f>=
\delta_{0,0}||\;|\tilde{f}>||_\sim^2$ !\\ 
Not only is the norm of an 
averaged vector with respect to the average inner product smaller than or 
equal to the norm of the original vector with respect to the original
inner product but also the average inner product and the original inner 
product in fact coincide on averaged vectors, nothing blows up.
The reason for this is that the spectrum is discrete so that generalized 
eigenfunctions are actually elements of the Hilbert space.

Let us return now to the constraint operator $\hat{C}'$ above and let 
us set $c=1$ for simplicity. We are actually able to describe explicitly
the complete spectrum of $\hat{C}'$, namely, it is the entire complex plane.
Generalized eigenvectors are given by $|z,z'>:=|z>\otimes |z'>$ where
$|z>$ are the usual coherent states and the eigenvalue is given by 
$z-z'$. Now, while the spectrum is not discrete just as in the case of
$\hat{H}=\hat{p}-\hat{q}$, still the generalized eigenvectors are proper,
normalized (although not orthonormal) elements of the Hilbert space just 
as in the case of $\hat{H}=\hat{a}^\dagger\hat{a}-\hat{b}^\dagger\hat{b}$ 
which is in contrast to the case $\hat{H}=\hat{p}-\hat{q}$. As we will see,
it is this difference that makes the group averaged solutions again 
normalizable with respect to the original inner product.\\ 
To see this we must first define a group averaging. To begin with notice that
all vectors of the form $|z,z>$ are annihilated by $\hat{C}'$. Moreover,
recall that coherent states are overcomplete but that still an identity
operator is given by 
$$
1_{{\cal H}}=\int \frac{d\bar{z} dz}{2i\pi} \int \frac{d\bar{z}' dz'}{2i\pi} 
|z,z'><z,z'|=:\int d\mu(\bar{z},z)\int d\mu(\bar{z}',z') |z,z'><z,z'|\;.
$$
Thus, in analogy with $\hat{H}=\hat{p}-\hat{q}$ we try to define an average 
map as follows : 
\ba \label{31}
|\tilde{f}>&:=&\hat{\eta}|f>:=\int \frac{dt}{2\pi} e^{it\hat{C}'}|f>
\nonumber\\
&=& \int d\mu(\bar{z},z)\int d\mu(\bar{z}',z') 
\int \frac{dt}{2\pi} e^{it\hat{C}'}|z,z'><z,z'|f>
\nonumber\\
&=& \int d\mu(\bar{z},z)\int d\mu(\bar{z}',z') \delta(z,z')
|z,z'><z,z'|f>
\ea
where $\delta(z):=\delta(x)_{x=z}$ for complex $z$ denotes the analytic 
continuation of the $\delta$ distribution. Interpreting this formal 
object as $\delta(x)\delta(y)$ with $z=x+iy$ results in 
\be \label{32}
|\tilde{f}>:= \int d\mu(\bar{z},z) |z,z> <z,z|f>
\ee
that is, $\hat{\eta}=\int d\mu(\bar{z},z) |z,z><z,z|$. 
Another way to justify (\ref{32}) is to interprete $\delta(z)$ as 
a holomorphic $\delta$ distribution in the sense $\int dz \delta(z,z_0) f(z)
=f(z_0)$ for holomorphic $f$ and to drop the remaining $d\bar{z}$ integral.
A reader who does 
not like the formal derivation (\ref{31}) may take (\ref{32}) as a definition
which is very natural and motivated by all the examples above. In fact,
apart from the explicit measure $d\mu(\bar{z},z)$, this $\hat{\eta}$ is 
the most general one that maps $|f>$ to a solution. Notice that $\hat{\eta}$
is the precise analogue of the average maps $\int dp |p,p><p,p|$ and 
$\sum_m |m,m><m,m|$ encountered above, all of them are projectors 
on the solution space and the measures $d\mu(\bar{z},z),dp,1$ are the 
diagonal measures that survive from the insertion of the $1_{{\cal H}}$
after evaluating the $\delta$ distribution.\\
It turns out that our $\eta$ can be written in a yet much simpler form in 
terms of the vectors $|N>$ above. First, by expanding the coherent 
states in terms of oscillator eigenfunctions, we observe that 
\ba \label{33}
|z,z>&=&e^{-|z|^2}\sum_{m,n=0}^\infty \frac{z^{m+n}}{\sqrt{m! n!}}|m,n>
\nonumber\\
&=& e^{-|z|^2}
\sum_{N=0}^\infty z^N\sum_{m=0}^N \frac{1}{\sqrt{m!(N-m)!}}|m,N-m>
\nonumber\\
&=&e^{-|z|^2} \sum_{N=0}^\infty \frac{z^N}{\sqrt{N!}} 2^{N/2} |N>  
\ea
where in the last step we have used definition (\ref{26}) of $|N>$.
Formula (\ref{33}) looks like the coherent state $|\sqrt{2}z>$ except that
the Fock states $|n>$ are replaced by the solution states $|N>$ of $\hat{C}'$
of level $N$. This confirms our earlier statement that every solution of
the constraint is a linear combination of solutions of the form $|N>$.\\
Next we explicitly perform the $\bar{z},z$ integrals involved in $\hat{\eta}$
and find (the calculation is exactly the same as the one for proving 
overcompleteness of coherent states)
\ba \label{34}
\hat{\eta}&=&\sum_{M,N=0}^\infty \frac{2^{(M+N)/2}}{\sqrt{M! N!}} |M><N|
\int d\mu(\bar{z},z) e^{-2|z|^2} z^M \bar{z}^N\nonumber\\
&=&\frac{1}{2}\sum_{M,N=0}^\infty \frac{1}{\sqrt{M! N!}} |M><N|
\int d\mu(\bar{z},z) e^{-|z|^2} z^M \bar{z}^N\nonumber\\
&=&\frac{1}{2}\sum_{M,N=0}^\infty \frac{1}{\sqrt{M! N!}} |M><N|
\delta_{M,N} N!\nonumber\\
&=& \frac{1}{2}\sum_{N=0}^\infty |N><N|\;.
\ea
Apart from the factor $1/2$ expression (\ref{34}) is the expected result :
the $\hat{\eta}$ operator is a projector on the solution space which can 
be more conveniently written in terms of the $|N>$'s. In the sequel we 
will disregard the $1/2$.\\ 
The group average inner product becomes 
\be \label{35}
<\tilde{f},\tilde{g}>_\sim=\sum_{N=0}^\infty <f,N><N,g>\;.
\ee
Notice that the $|N>$ are orthonormal among each other with respect to 
the original inner product. Let $\hat{\eta}'$ be the projector 
on the subspace of $\cal H$ orthogonal to the completion of the span of 
the $|N>$ so that $1_{{\cal H}}=\hat{\eta}+\hat{\eta}'$. Then, since 
both $\hat{\eta},\hat{\eta}'$  are positive operators we have
$$
||\;|\tilde{f}>||_\sim^2=<f,\hat{\eta}f>\le <f,f>=||\;|f>||^2$$ and 
$$
||\;|\tilde{f}>||^2=<f,\hat{\eta}^\dagger\hat{\eta} f>
=<f,\hat{\eta}\hat{\eta} f>
=<f,\hat{\eta} f>=||\;|\tilde{f}>||_\sim^2
$$
and we encounter the same phenomenon as for the constraint 
$\hat{H}:=\hat{a}^\dagger\hat{a}-\hat{b}^\dagger\hat{b}$ !\\
\\
What we learn from these model investigations is the following : 
\begin{itemize}
\item[1)] The assumption that averaged vectors are in general not 
normalizable with respect to the original inner product is false. 
Roughly, it is false for models in which generalized eigenvectors are in fact
elements of the Hilbert space. Examples include unbounded self-adjoint (with 
discrete spectrum) and non-self-adjoint constraint operators (with 
continuous spectrum) as we saw above. 
\item[2)] In the examples above we saw that the following chain of
inequalities holds 
\be \label{36}
||\;|\tilde{f}>||_\sim^2=||\;|\tilde{f}>||^2\le ||\;|f>||^2
\ee
that is, whenever $|f>$ has a norm, so does $|\tilde{f}>$ and moreover
we can compute it by means of the original or the group average inner 
product. Thus, the group average map can be extended from $\Phi$ to
all of $\cal H$.
\item[3)]
However, it is not true that 
$<\tilde{f},\tilde{g}>=<\tilde{f},\tilde{g}>_\sim$ for $f\not=g$. Thus,
the group average inner product is still needed and different from the 
original inner product. It is given by computing the inner products 
of the projections of general vectors into the solution space. The 
observation that $\hat{\eta}$ is a certain kind of projector is a general
observation and can serve as an abstract definition of the average map
for a general theory whose constraint operator is not self-adjoint.
\item[4)] One certainly can write down solutions to $\hat{C}'$ which are
well-defined distributions on $\Phi$, however, unless they are actually
elements of $\cal H$ they are not normalizable with respect to the
average inner product. All distributions which do not have a norm with 
respect to the original inner product have to be discarded.
\item[5)] The importance of coherent states for the solution of the 
quantum constraint lets us expect that coherent states will also play a 
quite important role in quantum gravity. Notice that coherent states are
{\em infinite} linear combinations of the basic states $|N>$ and so they 
are not cylindrical, however, they are elements of the Hilbert 
space while the states $|N>$ are in fact cylindrical.
\item[6)] We may view the definition
$$
\hat{\eta}:=\int_S d\nu(s) |s><s|,
$$
where $S$ is a complete set of labels of (generalized) solutions $|s>$ to a 
(not necessarily self-adjoint) dual constraint operator $\hat{C}'$ and 
$\nu$ is a ``natural" measure thereon, 
as a more general definiton of a group average map which in the case 
of self-adjoint $\hat{C}$ reduces to the usual definition. The measure
$\nu$, as we have seen, arises typically as follows : one considers 
first the spectral measure $\mu$ for the constraint operator $\hat{C}'$
and then $d\nu=\int_S  d\mu \delta(\hat{C}')$ is the measure induced by $\mu$
by deleting the integral over the spectrum $S$ of $\hat{C}'$. In case 
of a self-adjoint constraint operator this is precisely what happens
via the traditional group average approach.
\end{itemize}
Finally, let us address the question of how to model the {\em Lorentzian}
Hamiltonian constraint. Recall from \cite{1b} that the Lorentzian 
Hamiltonian constraint consists of two pieces 
$\hat{H}(N)=\hat{T}(N)-\hat{H}^E(N)$ where
$\hat{H}^E(N)$ is the Euclidean Hamiltonian constraint which can be modeled 
by a linear combination $\hat{a}^\dagger-\bar{c}\hat{b}$ as above. The 
piece $\hat{T}(N)$ on the other hand is a complicated multiple commutator 
between two factors of $\hat{H}^E(N)$ and three factors of the volume 
operator $\hat{V}$ and several holonomy operators along {\em open} paths. 
More precisely, it contains two factors of the 
form $h_s^{-1}[h_s,[\hat{V},\hat{H}^E(1)]]$ and one factor of the form 
$h_s^{-1}[h_s,\hat{V}]$ where $s$ is an open path. The operator $\hat{V}$ 
does not change the graph, it acts essentially by multiplication 
with functions that depend on the spin. The above combinations of holonomies 
along open paths and back also do not change the graph. Thus it is only the 
two factors of 
$\hat{H}^E(N)$ that change the graph. Moreover, since every commutator 
decreases the spin power by one unit we conclude that the spin power
of $h_s^{-1}[h_s,[\hat{V},\hat{H}^E(1)]]$ is of order zero (recall that 
the spin power of $\hat{H}^E(N)$ is $1/2$ to see this). Thus the spin 
power of 
$\hat{T}(N)$ is $1/2$ as well. Since the ``spin" power of $\hat{a}^\dagger$
is $1/2$ as well we conclude that the bilinear operators 
$(\hat{a}^\dagger)^2,(\hat{b}^\dagger)^2,\hat{a}^\dagger\hat{b}^\dagger$
appropriately model $\hat{T}(N)$ : they create {\em two} new excitations 
and their spin power is even higher than that of $\hat{T}(N)$. In order 
to account for the difference in the spin power one could choose to 
divide by a square root of the number operator but we will not do that 
because that wouuld pushh the analogy too far and even if we would do
it the argument displayed below would still apply. We 
therefore propose to study the constraint operator 
$$
\hat{C}'=\alpha(\hat{a}^2+\hat{b}^2)+\beta\hat{a}\hat{b}
-\gamma(\hat{a}+\hat{b})
$$ 
where the choice of coefficients respects the fact that the coordinates $x,y$
have to be treated on equal footing in order to mirror the expression of 
$\hat{T}(N)$.\\
We can immediately solve $\hat{C}'$ : consider a state $|z,z'>$ as above,
then $\hat{C}'|z,z'>=[\alpha(z^2+(z')^2)+\beta z z'-\gamma(z+z')]|z,z'>$
is an eigenfunction of $\hat{C}'$, again the operator $\hat{C}'$ is 
diagonalized by proper elements of $\cal H$ ! The condition that the 
eigenvalue vanishes leads to a condition $z'=h(z)$ where $h(z)$ 
is a holomorphic function of $z$ (possibly on a two-sheeted 
non-compact Riemann surface depending on the choice of the parameters 
$\alpha,\beta,\gamma$). The same arguments as above now suggest to define the
average operator by
\be \label{37}
\hat{\eta}:=\int d\mu(\bar{z},z) |z,h(z)><z,h(z)|\;
\ee
Expression (\ref{37}) is sufficient to prove that for any cylindrical
state $|f>$ (that is, they are finite linear combinations of the states 
$|m,n>$) the averaged state $|\tilde{f}>:=\hat{\eta}|f>$ is still an 
element of the Hilbert space. Namely, since $||\;|z,h(z)>||=1$
we have by the Schwarz inequality
\ba \label{38}
&&||\;|\tilde{f}>||^2\nonumber\\
&\le& \int d\mu(\bar{z},z) |<f|z,h(z)>|
\int d\mu(\bar{z}',z') |<f|z',h(z')>|\; |<z,h(z)|z',h(z')>|\nonumber\\
&\le& (\int d\mu(\bar{z},z) |<f|z,h(z)>|)^2\;.
\ea
But $<f|z,h(z)>=p(z,h(z))e^{-[|z|^2+|h(z)|^2]}$ where $p(z,h(z))$ is a 
polynomial depending on $|f>$ so that (\ref{38}) obviously converges.
We also have
\be \label{39}
<\tilde{f},\tilde{g}>_\sim=<f,\hat{\eta}g> \mbox{ and }
||\;|\tilde{f}>||_\sim^2=
\int d\mu(\bar{z},z) |<f|z,h(z)>|^2 <\infty
\ee
for the same reason. This time we cannot find an obvious relation between
$||\;|\tilde{f}>||_\sim,||\;|\tilde{f}>||,||\;|f>||$ though.\\
We conclude that the normalizability of averaged vectors with respect to the 
original inner product carries over to the more complicated ``Lorentzian"
constraint.

\subsection{A physical inner product for quantum general relativity}

Let us first check whether the inner product that results from averaging 
the diffeomorphism constraint actually is of the projector type mentioned 
above. This is easily seen to be the case : if we write
$\hat{\eta}_{Diff}:=\sum_{[T^S_I]} [T^S_I][T^S_I]^\dagger$ then
we recover indeed $\hat{\eta}_{Diff}(T^S_I)=[T^S_I]$, i.e. we have 
demonstrated that $\hat{\eta}_{Diff}$ is of the general type encountered 
in the examples, namely it is a projector formed by elementary solutions 
which themselves are averages of a suitable basis of the original Hilbert 
space.

Next, let us recall the structure of the solution space of Lorentzian 
quantum gravity from \cite{1c}.
Consider all possible labels $I$ of spin-network states
and call the resulting set $W$. The set $W$ splits naturally into two
parts : a piece $W_0$, called the set of {\em sources} and its complement 
$\overline{W}_0$
in $W$. The set $W_0$ is characterized by the property that no vector $T_I$
for $I\in W_0$ can occur in the (decomposition into spin-networks of the)
image of the Euclidean Hamiltonian constraint $\hat{H}^E(N)$ on $\Phi$ 
for no choice of 
prescription $p$. The set $\overline{W}_0$ therefore contains the labels which
are sufficient to span the image of $\hat{H}^E(N)$. It can be further 
decomposed as follows : for each $I\in W_0$ let $W^{(n)}(I)$ be the set 
of labels that one finds by decomposing $[\hat{H}^E(N)]^n T_I$ into 
spin-network states for all possible $N$ and for all possible prescriptions
$p$. These are the states of level $n$ with source $I$ and they are the 
precise analogue of the states $|m,N-m>$ of level $N$ considered above.
We now take the group average with respect to the diffeomorphism constraint
and it then turns out that all the states with labels $[J]$ coming from
different $W^{(n)}(I)$ (up to a diffeomorphism) are orthonormal. Thus the 
diffeomorphism invariant Hilbert space splits as 
$$
{\cal H}_{Diff}=\oplus_{[I],I\in W_0} {\cal H}_{[I]}
\mbox{ where }
{\cal H}_{[I]}=\oplus_{n=0}^\infty {\cal H}_{[W^{(n)}(I)]}
$$
and where $[W^{(n)}(I)]$ denotes the group averaged labels of
$W^{(n)}(I)$. The space ${\cal H}_{[W^{(n)}(I)]}$
is obviously a finite dimensional vector space and so 
${\cal H}_{[I]}$ is separable. Moreover, $\hat{H}'(N)$ maps 
${\cal H}_{[I]}$ into itself, more precisely, it maps 
${\cal H}_{[W^{(n)}(I)]}$ into 
${\cal H}_{[W^{(n-1)}(I)]}\cup{\cal H}_{[W^{(n-2)}(I)]}$. Therefore 
the Hamiltonian constraint can be solved on each ${\cal H}_{[I]}$
separately. Let us label the vectors of ${\cal H}_{[W^{(n)}(I)]}$ 
by $T^n_i$ (notice that all these vectors are diffeomorphism invariant).
Then there are matrices $a^n_{ij},b^n_{ij}$ which depend on $[I]$ and on 
one of the (set of diffeomorphic) vertices $v$ of the graph underlying $I$
such that $\hat{H}'_v T^n_i=a^n_{ij} T^{n-1}_j-b^n_{ij} T^{n-2}_j$. 
Writing a general element of $\Phi'$ as a linear combination of 
distributions of the form (one for each $[I]$)
$\sum_n c^n_i T^n_i$ we find that 
$c^{n+1}_i a^{n+1}_{ij}=c^{n+2}_i b^{n+2}_{ij}$ for each $[I],n,j,v$.
It turns out that the matrices $a^n,b^n$ are sufficiently degenerate 
as to allow for solutions $c^n_i$ such that 
$c^{n+1}_i a^{n+1}_{ij}=c^{n+2}_i b^{n+2}_{ij}=0$ for each $[I],n,j,v$
and so does not mix levels but a general solution will mix levels.\\
The precise growth of the coefficients of the matrices $a^n_{ij},b^n_{ij}$
is unknown (although the reasoning from above suggests that they do not 
grow worse than $j^{1/2}$ where $j$ is the total spin of a state). On the 
other hand, if $b$ was identically zero as 
it is the case for the Euclidean Hamiltonian constraint then levels 
do not get mixed and then the example $\hat{C}'=\hat{a}-\hat{b}$ considered 
above suggests (although we have no proof at present, of course)
that a general solution will be normalizable with respect to the inner 
product on ${\cal H}_{Diff}$. The example
$\hat{C}'=\alpha(\hat{a}^2+\hat{b}^2+\beta\hat{a}\hat{b}
-\gamma(\hat{a}+\hat{b})$ studied above suggests that the same is true 
for the full Lorentzian constraint, i.e. $b\not= 0$ (again we have no proof 
for this).\\ 
Suppose now that we have explicitly solved the infinite 
number of {\em linear} equations 
$c^{n+1}_i a^{n+1}_{ij}=c^{n+2}_i b^{n+2}_{ij}$ for each $[I],n,j,v$
exactly (this is actually possible, although quite complicated; we expect 
that one can invent some efficient code that lets us write down the 
solutions at least up to a controllable error). The examples studied
above suggest that every solution is a kind of coherent state. 
The first possibility is that every solution lies in ${\cal H}_{Diff}$.
Then, since
${\cal H}_{[I]}$ is separable, one can in principle 
find an orthonormal basis $\{T_{[I],\mu}\}_{\mu=0}^\infty$ (with respect 
to $<.,.>_{Diff}$) on the solution 
space for each $[I]$ which defines actually a subspace of ${\cal H}_{[I]}$.
If not every solution lies in ${\cal H}_{Diff}$ then we can still find a 
countable generalized basis $\{T_{[I],\mu}\}_{\mu=0}^\infty$ built from 
infinite linear combinations of elements of ${\cal H}_{[I]}$ which are 
orthonormal
(with respect to $<.,.>_{Diff}$) in the sense of generalized eigenvectors,
that is, $<T_{[I],\mu},T_{[J],\nu}>_{Diff}=0$ unless $[I]=[J]$ and 
$\mu=\nu$ in which case this quantity diverges. (This is quite similar to
the momentum eigenfunction ``normalization" in the sense of 
$\delta$ distributions : $<p,p'>=\delta(p,p')$. It is also precisely
what happens with the group averaged spin-network states : with respect
to the inner product on $\cal H$ they are orthogonal in the sense that
$$
<[T_I],[T_J]>:=\sum_{T\in\{T_J\}} [T_I](T)=0
$$
unless $[I]=[J]$ in which case this quantity blows up). \\
In either case the group average proposal then leads us to define the 
projectors 
\be \label{40}
\hat{\eta}_{Ham}:=\sum_{[I]} \hat{\eta}_{[I]} \mbox{ where }
\hat{\eta}_{[I]}=\sum_{\mu=0}^\infty T_{[I],\mu} T_{[I],\mu}^\dagger,
\ee
physical states as arising from group averaging are defined
by $T_{phys}:=\hat{\eta} T$ for each $T\in \Phi'_{Diff}$ 
and the physical inner product between such physical states is given by
\be \label{41}
<T_{phys},T'_{phys}>_{phys}:=<T,\hat{\eta} T'>_{Diff}=
T_{phys}(T')
\ee
where the latter notation means evaluation of the distribution 
$T_{phys}\in(\Phi'_{Diff})'$ on the ``test function" $T'\in\Phi'_{Diff}$ by 
means of the inner product of ${\cal H}_{Diff}$.\\
Equation (\ref{41}) is a physical inner product for Lorentzian quantum
general relativity which is naturally suggested to us by group average 
kind of reasonings. It is precisely of the structure of the 
group average map for the diffeomorphism constraint $\eta_{Diff}$ 
displayed at the beginning of this section.\\
Remarks :
\begin{itemize}
\item[1)] We want to stress that the inner product (\ref{41}) exists
even if the solutions $T_{[I],\mu}$ are not
normalizable with respect to $<.,.>_{Diff}$ although we expect this to 
be true. We have struggled to 
list arguments in favour of normalizabilty of $T_{[I],\mu}$ only in order to 
remove the feeling of discomfort : The quantum mechanics example  
$\hat{H}=\hat{a}^\dagger\hat{a}-\hat{b}^\dagger\hat{b}$ shows that
physical states as obtained by rigorous group averaging can be normalizable 
with respect to the original inner product without that anything is wrong 
with that. 
\item[2)] The fact that solutions to the Hamiltonian constraint 
are presumably normalizable 
with respect to the diffeomorphism invariant inner product can be 
interpreted as saying that the whole distributionality of a solution to 
{\em all} constraints is already captured by averaging with respect to 
the diffeomorphism constraint alone. In fact, if we had not split the 
solution of diffeomorphism and Hamiltonian constraint into two steps but 
had solved all constraints in one stroke without even constructing
$\eta_{Diff}$ then we would not have even noticed this fact.
\item[3)] As one can show, the ``flat state" $\delta(F)$ can be given 
rigorous meaning as a distribution on $\Phi'_{Diff}$ \cite{7} and it solves
$\hat{H}^E(N)$ at least. In 2+1 Euclidean gravity the flat state is a 
physical state for the curvature constraint $F=0$. On the other hand,
to make 2+1 Euclidean gravity resemble 3+1 Lorentzian gravity one can replace
$F=0$ by Diffeomorphism and Hamiltonian constraints \cite{3}. One finds
that in 2+1 it is indeed true that physical solutions as slected by the so 
defined Hamiltonian constraint (analogously constructed as in this paper
for 3+1 Lorentzian gravity) are normalizable with respect to $<.,.>_{Diff}$.
It follows then that the flat state $\delta(F)$ is {\em 
not} normalizable with respect to $<.,.>_{phys}$ because it is an 
uncountably infinite linear combination of diffeomorphism invariant 
spin-network states with all coeeficients of order unity \cite{3} and so is 
not even normalizable with respect to $<.,.>_{Diff}$. Thus, the inner products 
of a topological quantum field theory (like 2+1 Euclidean gravity in the 
Witten formulation) and quantum general relativity in 3+1 dimensions 
have inner products which are 
presumably very singular with respect to each other as one would expect.
To prevent wrong conclusions, notice 
that $F=0$, which can be solved, for instance, by $A=0$, does not mean that
we are in Minkowski space : $A=\Gamma+K=0$ just does not have any obvious
Minkowski space interpretation.
\item[4)] Notice that since the projector inner product (\ref{41}) is 
induced from that of $\cal H$ for which the classical reality conditions 
are implemented as adjointness relations, we can claim that (\ref{41})
incorporates the {\em physical} reality conditions on Dirac observables
provided a Dirac observable is self-adjoint on $\cal H$ already.
This is because a Dirac observable leaves the solution space invariant
and is therefore already projected (``it commutes with the group 
averaging"), see below.
\item[5)] A complete set of Dirac observables in the sense that it be a 
self-adjoint and densely defined operator on ${\cal H}_{phys}$ are trivial
construct : Define 
\be \label{42}
\hat{O}'_{[I]\mu,[J]\nu}:=T_{[I]\mu}(T_{[J]\nu})^\dagger \mbox{ and }
\hat{O}_{[I]\mu,[J]\nu}:=
\frac{\hat{O}'_{[I]\mu,[J]\nu}+\hat{O}'_{[I]\mu,[J]\nu}}{2}
\ee
where the dagger operation is with respect to $<.,.>_{phys}$
then linear combinations $\hat{O}$ of symmetric operators of the form 
(\ref{42}) form a complete set of Dirac observables (modulo domain 
questions).\\
Notice that in case that the $T_{[I]\mu}$ are not $<.,.>_{Diff}$ normalizable
then (\ref{42}) has to be understood in the folowing sense : there are
pre-images $f_{[I]\mu}\in\Phi$ such that $\hat{\eta} 
f_{[I]\mu}=T_{[I]\mu}$. Now the $T_{[I]\mu}$ become normalizable with respect
to $<.,.>_{phys}$ because by definition 
$<T_{[I]\mu},T_{[J]\nu}>_{phys}:=T_{[I]\mu}(f_{[J]\nu})$. In case of general
relativity the $f_{[I]\mu}$ are related to the ``source states" 
constructed in \cite{1c} which are really elements of $\Phi$ and 
orthonormal with respect
to $<.,.>$ (after diffeomorphism group averaging they form an orthonormal 
system (but not a basis) contained in ${\cal H}_{[I]}$). This implies
that the $T_{[I]\mu}$ are orthonormal with respect to $<.,.>_{phys}$.\\
From this follows then immediately that 
$\hat{\eta}\hat{O}=\hat{O}\hat{\eta}$. Also, $\hat{O}$ is self-adjoint
on $\cal H$ if we define 
$<f,\hat{O}'_{[I]\mu,[J]\nu}g>:=\overline{T_{[I]\mu}(f)}
T_{[J]\nu}(g)$.\\
\\
\end{itemize}

{\large Acknowledgements}\\
\\
Many of the issues discussed in this article were raised in the course
of inspriring discussions with Abhay Ashtekar, Jurek Lewandowski, Don 
Marolf and Jos\'e Mour\~ao during the wonderful ESI workshop in Vienna,
July-August 1996. The author would like to thank the ESI and the 
organizers of 
the workshop, Peter Aichelburg and Abhay Ashtekar, for providing this 
excellent research environment.\\
This research project was 
supported in part by the ESI and by DOE-Grant DE-FG02-94ER25228 to Harvard 
University.


\begin{thebibliography}{99}

\parskip -5pt

\bibitem{1a} T. Thiemann, Physics Letters B {\bf 380} (1996) 257-264 

\bibitem{1b} T. Thiemann, ``Quantum Spin Dynamics (QSD)"
Harvard University Preprint HUTMP-96/B-359, gr-qc/9606089

\bibitem{1c} T. Thiemann, ``Quantum Spin Dynamics (QSD) II : The Kernel
of the Wheeler-DeWitt Constraint Operator"
Harvard University Preprint HUTMP-96/B-352, gr-qc/9606090

\bibitem{Baez} J. Baez, contribution to the newsletter ``Matter of 
Gravity", webpage html://www.phys.psu.edu/PULLIN/, gr-qc/9609008

\bibitem{AI} A. Ashtekar and C.J. Isham,
Class. \& Quan. Grav. {\bf 9}, 1433 (1992).

\bibitem{AL1} A. Ashtekar and J. Lewandowski, ``Representation
theory of analytic holonomy $C^\star$ algebras'', in {\it Knots and
quantum gravity}, J. Baez (ed), (Oxford University Press, Oxford 1994).

\bibitem{AL2} A. Ashtekar and J. Lewandowski, ``Differential
geometry on the space of connections via graphs and projective
limits'', Journ. Geo. Physics {\bf 17} (1995) 191

\bibitem{AL3} A. Ashtekar and J. Lewandowski, J. Math. Phys. {\bf 36}, 2170
(1995). 

\bibitem{MM} D. Marolf and J. M. Mour\~ao, ``On the support of the
Ashtekar-Lewandowski measure'',  Commun. Math. Phys. {\bf 170} (1995)
583-606

\bibitem{almmt} A. Ashtekar, J. Lewandowski, D. Marolf, J. Mour\~ao, T.
Thiemann, Journ. Math. Phys. {\bf 36} (1995) 519-551

\bibitem{AA} A.\ Ashtekar, Phys.\ Rev. Lett.\ {\bf 57} 2244 (1986),
            Phys.\ Rev.\ {\bf D36}, 1587 (1987).

\bibitem{Barbero} F. Barbero, Phys. Rev. D {\bf 51} (1995) 5507


\bibitem{Wick} A.\ Ashtekar, J.\ Lewandowski, D.\ Marolf, J.\ Mour\~ao, T.\
            Thiemann, J. Funct. Analysis {\bf 135} (1996) 519-551\\
            T. Thiemann, Class. Quantum Gravity {\bf 13} (1996) 1383-1403\\
            A. Ashtekar, Phys. Rev. {\bf D53} (Rapid Communications)
            R2865-69 (1996)

\bibitem{Gambini} R. Gambini, A. Trias, Nucl. Phys. {\bf B22} 1380 (1980) 

\bibitem{RS1} C. Rovelli, L. Smolin, Nucl. Phys. {\bf B331}, 80 (1990). 

\bibitem{RR} M. Reisenberger, C. Rovelli,  ``Sum over surfaces" form of Loop 
Quantum Gravity, gr-qc/9612035

\bibitem{K} K. Krasnov, Phys. Rev. {\bf D55} 3505 (1997)

\bibitem{R} C. Rovelli, Phys. Rev. Lett. {\bf 77} 3288 (1996)

\bibitem{AK} A. Ashtekar, K. Krasnov (in preparation)

\bibitem{2} T. Thiemann,``QSD V : Quantum Gravity as the natural 
Regulator of Matter Quantum Field Theories", Harvard University 
Preprint HUTMP-96/B-357

\bibitem{2a} T. Thiemann,``QSD VI : 
Quantum Poincar\'e Algebra and a Quantum Positivity of Energy Theorem
for Canonical Quantum Gravity", Harvard University Preprint HUTMP-97/B-356

\bibitem{3} T. Thiemann, ``QSD IV : 2+1 Euclidean Quantum Gravity as a 
model to test 3+1 Lorentzian Quantum Gravity"", HUTMP-96/B-360

\bibitem{5} A. Higuchi Class. Quant. Grav. {\bf 8},  1983 (1991)\\
            A. Higuchi Class. Quant. Grav. {\bf 8}, 2023 (1991)

\bibitem{4} D. Marolf, ``The spectral analysis inner product for
quantum gravity,'' preprint gr-qc/9409036, to appear in the
Proceedings of the VIIth Marcel-Grossman Conference, R. Ruffini and
M. Keiser (eds) (World Scientific, Singapore, 1995); D. Marolf,
Ph.D. Dissertation, The University of Texas at Austin (1992)\\
D. Marolf, ``Quantum observable and recollapsing
dynamics,'' preprint gr-qc/9404053. Class. Quant. Grav. {\bf 12} (1995) 
1199\\
D. Marolf ``Almost Ideal Clocks in Quantum Cosmology: A Brief
Derivation of Time,'' preprint gr-qc/9412016.

\bibitem{TTCoh} T. Thiemann, ``Coherent States for Quantum General 
Relativity" (in preparation)

\bibitem{BaSa} J. Baez, S. Sawin, ``Functional Integration on Spaces
of Connections", q-alg/9507023

\bibitem{TTILT} T. Thiemann, ``The Inverse Loop Transform", 
Harvard University Preprint HUTMP-95/B-346,
hep-th/9601105

\bibitem{LM} J. Lewandowski, D. Marolf (in preparation)

\bibitem{Z} J.-A. Zapata, ``A combinatoric approach to diffeomorphism
invariant Quantum Gauge Theories", gr-qc/9703037\\
J.-A. Zapata, ``A combinatoric space from loop quantum gravity", 
gr-qc/9703038

\bibitem{TTL} T. Thiemann, ``A length operator for canonical quantum 
gravity", Harvard University Preprint HUTMP-96/B-354, gr-qc/9606092

\bibitem{Geometry} C. Rovelli, L. Smolin, Nucl. Phys. {\bf B442} (1995) 953,
Erratum : Nucl. Phys. {\bf B456} (1995) 734\\
A. Ashtekar, J. Lewandowski, Class. Quantum Grav. {\bf 14} (1997) A55-81\\
J. Lewandowski, Class. Quantum Gravity {\bf 14}  (1997) 71-76

\bibitem{6} P. Haj\'i\v{c}ek, K. Kucha\v{r}, Phys. Rev. {\bf D41}
(1990) 1091, Journ. Math. Phys. {\bf 31} (1990) 1723

\bibitem{7} D. Marolf, J. Mour\~ao, T. Thiemann, 
``The Status of Diffeomorphism Superselection in Euclidean 2+1 Gravity",
HUTMP-97/B-360, gr-qc/9701068


\end{thebibliography}
\end{document}